\documentclass[apj,twocolumn,twocolappendix,numberedappendix]{openjournal}
\usepackage[dvipsnames, table]{xcolor}
\usepackage{newtxtext,newtxmath,enumitem,multirow}
\usepackage[utf8]{inputenc}
\usepackage[T1]{fontenc}
\usepackage[breaklinks,colorlinks,allcolors=blue,citecolor=blue,urlcolor=blue]{hyperref}
\usepackage{orcidlink}
\usepackage{graphicx}	
\usepackage{amsmath}	
\usepackage{amssymb}	
\usepackage{pifont}
\usepackage{mathrsfs}
\usepackage{dsfont}
\usepackage{blindtext}
\usepackage{nicematrix}
\usepackage{array, color, colortbl}
\usepackage{arydshln}
\usepackage{booktabs}
\usepackage{tabularx}

\pdfoutput=1

\newcommand{\LCDM}{$\Lambda$CDM\ }

\newcommand{\cMpc}{\: \mathrm{cMpc}}
\newcommand{\hMpc}{\: h^{-1} \, \mathrm{cMpc}}
\newcommand{\Msun}{\: \mathrm{M}_{\odot}}
\newcommand{\K}{\: \mathrm{K}}

\shorttitle{Interpretable ML of halo gas density profiles: a sensitivity analysis of cosmological hydrodynamical simulations}
\shortauthors{D. Sorini et al.}

\begin{document}
\title[Interpretable machine learning of halo gas density profiles: a sensitivity analysis of cosmological hydrodynamical simulations]{Interpretable machine learning of halo gas density profiles: a sensitivity analysis of cosmological hydrodynamical simulations}
\author{Daniele Sorini\,\orcidlink{0000-0001-5252-9561}$^{1,\star}$}
\author{Sownak Bose\,\orcidlink{0000-0002-0974-5266}$^{1}$}
\author{Mathilda Denison\,\orcidlink{0009-0007-1790-7337}$^{2,1}$}
\author{Romeel Dav\'e\,\orcidlink{0000-0003-2842-9434}$^{3,4,5}$}
\affiliation{$^{1}$Institute for Computational Cosmology, Department of Physics, Durham University, South Road, Durham, DH1 3LE, United Kingdom}
\affiliation{$^{2}$Department of Physics and Astronomy, University of Pennsylvania, 209 South 33rd Street, Philadelphia, PA 19104-6396, USA}
\affiliation{$^{3}$Institute for Astronomy, University of Edinburgh, Royal Observatory, Blackford Hill, Edinburgh, EH9 3HJ, United Kingdom}
\affiliation{$^{4}$Department of Physics and Astronomy, University of the Western Cape, Bellville, Cape Town 7535, South Africa}
\affiliation{$^{5}$South African Astronomical Observatory, Observatory, Cape Town, 7925, South Africa}
\thanks{$^\star$\href{mailto:daniele.sorini@durham.ac.uk}{daniele.sorini@durham.ac.uk}}

\begin{abstract}
Stellar and AGN-driven feedback processes affect the distribution of gas on a wide range of scales, from within galaxies well into the intergalactic medium. Yet, it remains unclear how feedback, through its connection to key galaxy properties, shapes the radial gas density profile in the host halo. We tackle this question using suites of the EAGLE, IllustrisTNG, and Simba cosmological hydrodynamical simulations, which span a variety of feedback models. We develop a random forest algorithm that predicts the radial gas density profile within haloes from the total halo mass and five global properties of the central galaxy: gas and stellar mass; star formation rate; mass and accretion rate of the central black hole (BH). The algorithm reproduces the simulated gas density profiles with an average accuracy of $\sim$83--90\% over the halo mass range $10^{9.5} \, \mathrm{M}_{\odot} < M_{\rm 200c} < 10^{15} \Msun$ and redshift interval $0<z<4$. For the first time, we apply Sobol statistical sensitivity analysis to full cosmological hydrodynamical simulations, quantifying how each feature affects the gas density as a function of distance from the halo centre. Across all simulations and redshifts, the total halo mass and the gas mass of the central galaxy are the most strongly tied to the halo gas distribution, while stellar and BH properties are generally less informative. The exact relative importance of the different features depends on the feedback scenario and redshift. Our framework can be readily embedded in semi-analytic models of galaxy formation to incorporate halo gas density profiles consistent with different hydrodynamical simulations. Our work also provides a proof of concept for constraining feedback models with future observations of galaxy properties and of the surrounding gas distribution.
\end{abstract}


\maketitle


\section{Introduction}
\label{sec:intro}

Understanding how gas is distributed within galaxies and throughout the large-scale structure of the Universe is key to developing a comprehensive theory of galaxy formation in a cosmological context. The efficiency with which galaxies form stars depends critically on the amount, spatial configuration, and thermodynamic state of gas, all of which are regulated by stellar and AGN-driven feedback processes. Despite extensive theoretical and observational efforts, these processes remain poorly constrained and continue to represent one of the main challenges in galaxy formation theory \citep[see e.g.][]{feedback_review, Crain_2023}.  

A wealth of observational campaigns has been devoted to mapping the gaseous environments of galaxies, both in absorption and emission. Early efforts focused on Lyman-$\alpha$ absorption in the circumgalactic medium (CGM) of high-redshift galaxies \citep[e.g.][]{Steidel_2010, Rudie_2012_CGM} and quasars \citep[e.g.][]{QPQ2, Prochaska_2013}, while subsequent studies extended to lower redshifts and incorporated metal-line diagnostics \citep[e.g.][]{Farina_2014, Johnson_2015}. In parallel, advances in integral-field spectroscopy, notably with MUSE \citep{MUSE}, have enabled the detection of extended Lyman-$\alpha$ and metal-line emission around galaxies and quasars over a broad redshift range \citep[e.g.][]{Arrigoni-Battaia_2019, Galbiati_2024}. More recently, X-ray and millimetre-wave observations have provided complementary constraints on the hot and ionised phases of halo gas through measurements of X-ray surface brightness \citep{Lyskova_2023, Zhang_2024} and the thermal and kinetic Sunyaev-Zeldovich effects \citep[e.g.][]{Amodeo_2021}. The convergence of multi-wavelength datasets now allows the gas distribution to be traced across a wide mass and redshift interval, probing different phases of the baryonic ecosystem. Comparing these observations with theoretical predictions is essential for constraining models of feedback and galaxy evolution.  

On the theoretical side, analytic \citep{HS03, SP21, Pandya_2023, Sorini_2024_Lambda} and semi-analytic \citep{Cole_1994, Kauffmann_1999, Lacey_2016} models of galaxy formation have often assumed spherically symmetric power-law density profiles. This assumption was later supported by cosmological simulations, particularly in the outskirts of haloes \citep[e.g.][]{Stinson_2015}. Nevertheless, analytic and semi-analytic approaches cannot capture the complex interplay of feedback mechanisms that regulate the gas distribution. Cosmological hydrodynamical simulations provide a more complete description by self-consistently modelling gas dynamics and the exchange of energy and momentum between baryons and dark matter (DM). 

However, different codes adopt distinct sub-grid prescriptions for stellar and AGN feedback, leading to markedly different predictions for the gas density profiles and thermodynamic structure of the CGM and intracluster medium \citep[e.g.][]{Pallottini_2014, Sorini_2020, Li_2023}. Studies based on the CAMELS suite of simulations \citep{CAMELS} have shown that variations in feedback parameters strongly affect the spread of gas on both small and large scales \citep{Gebhardt_2024}. Likewise, comparisons among the EAGLE \citep{EAGLE_Schaye2015}, IllustrisTNG \citep{IllustrisTNG}, and Simba \citep{Simba} simulations reveal significant model-dependent differences in the predicted gas distribution within and outside haloes \citep{Borrow_2020, Sorini_2022, Ayromlou_2023}. These findings highlight the need for robust frameworks to interpret how feedback physics shapes the baryon content of haloes and to test such frameworks against simulation suites with differing feedback implementations.

In this context, \cite{Sorini_2024} investigated the gas density profile around haloes in several variants of the Simba simulation incorporating different feedback prescriptions. In all cases, they found that the profiles can be well approximated with a universal analytical fitting function that depends on the total halo mass and redshift. However, the gas distribution is unlikely to depend on these two parameters only. In fact, the distribution and physical state of the CGM is the result of the complex interplay of several astrophysical processes, from star formation to stellar and AGN feedback \citep{Sorini_2022, Appleby_2023}. Such high complexity is hard to capture with simple analytical models.

Conversely, data-driven approaches excel at discovering intricate patterns linking several properties. For example, \cite{Appleby_2023_ML} applied a machine-learning (ML) algorithm on the Simba simulation, successfully predicting the thermal state of the CGM from mock quasar absorption line observables. This kind of approach is certainly promising for constraining galaxy formation models from observations. However, ML models are occasionally criticised due to perceived lack of explanatory power \citep[e.g.][]{Ntampaka_2011}. Nevertheless, this need not be the case: there are mathematical tools that can greatly aid the physical interpretation of a data-driven model, providing valuable insights.

In this work, we combine the strengths of ML and statistical sensitivity analysis \citep{Fisher_1918, Sobol_1993} to develop an accurate and physically interpretable data-driven model to to understand how different feedback scenarios impact galaxy properties and the gas distribution within the host halo. We train our model on data extracted from cosmological hydrodynamical simulations that incorporate distinct sub-grid physics: EAGLE, IllustrisTNG, and several variants of the Simba simulations. We test the predictions of our model over a total halo mass range of $10^{9.5} - 10^{15} \Msun$ and redshift interval $0<z<4$, obtaining an accuracy of $\sim 84-90\%$ above $z=1$ and $\sim 80-87\%$ at $z=0$, depending on the simulation. This level of accuracy is promising for future comparisons with the gas density profiles deduced from current absorption-line and emission-line observations in different wavelength bands \citep{Prochaska_2013, Zhang_2024, Sorini_2024_proc}. Thanks to sensitivity analysis, we are able to identify the galaxy and halo properties that have the strongest impact on the gas density at different distances from the halo centre.

This manuscript is organised as follows. In \S~\ref{sec:simulations} we briefly describe the main characteristics of the simulations utilised in this work. In \S~\ref{sec:model} we explain how we extracted the gas density profiles and galaxy properties from the simulations, and the details of our ML algorithm. The resulting data-driven model is then tested against the simulation data in \S~\ref{sec:validation}, where we quantify its accuracy. We then apply sensitivity analysis to physically interpret our model: \S~\ref{sec:sensitivity} explains the basics of this technique, while \S~\ref{sec:discussion} discusses the physical implications of the results. We summarise conclusions and perspectives of our work in \S~\ref{sec:conclusions}.

Throughout this paper, unless otherwise stated, we denote comoving lengths by prepending a `c' to the relevant unit (e.g., ckpc, cMpc, etc.). Without the `c' prefix, units are understood to be proper lengths. 

\section{Simulations}
\label{sec:simulations}

\setlength{\tabcolsep}{0pt}         
\renewcommand{\arraystretch}{1.4}   
\begin{table*}
    \caption{Properties of the simulations utilised in this work.}
    \label{tab:simulations}
    \begin{tabularx}{0.98\textwidth}{@{}c@{}c@{}c@{}l@{}c@{}c@{}c@{}l@{}}
    \toprule
        Project & Simulation & Box size & $\quad N_{\mathrm{res. el.}} \quad$ & $\quad m_{\mathrm{DM}} \quad$ & $\quad m_{\mathrm{gas}} \quad$ & $\varepsilon_{\rm gas}$ & $\quad$Feedback modes \\
         &  & [$\rm cMpc$] & & $\quad[\rm M_{\odot}]\quad$ & $[\rm M_{\odot}]$ & $[\rm pc]$ & \\
        \toprule
        
        EAGLE & \textbf{EAGLE-100} & \textbf{100} & $\quad \mathbf{2 \times 1504^3}\quad$ & $\quad \mathbf{9.7 \times 10^5} \quad$ & $\quad \mathbf{1.81 \times 10^6} \quad$ & \textbf{700} & $\quad$Stellar feedback, thermal AGN feedback \\
        
         & \cellcolor{gray!20} EAGLE-50 & \cellcolor{gray!20} 50 & \cellcolor{gray!20} $\quad 2 \times 752^3 \quad$ & \cellcolor{gray!20} $\quad 9.7 \times 10^6 \quad$ & \cellcolor{gray!20} $\quad 1.81 \times 10^6 \quad$ & \cellcolor{gray!20} 700 &  \cellcolor{gray!20} \\
         
         & \cellcolor{gray!20} EAGLE-25 & \cellcolor{gray!20} 25 & \cellcolor{gray!20} $\quad 2 \times 376^3 \quad$ & \cellcolor{gray!20} $\quad 9.7 \times 10^6 \quad$ & \cellcolor{gray!20} $\quad 1.81 \times 10^6 \quad$ & \cellcolor{gray!20} 700 & \cellcolor{gray!20} \\
         
         & \cellcolor{gray!20} EAGLE-LowRes & \cellcolor{gray!20} 100 & \cellcolor{gray!20} $\quad 2 \times 752^3 \quad$ & \cellcolor{gray!20} $\quad 7.76 \times 10^7 \quad$ & \cellcolor{gray!20} $\quad 1.45 \times 10^7 \quad$ & \cellcolor{gray!20} $\quad$1400$\quad$ &  \cellcolor{gray!20} \\
         
        \midrule
        
        $\quad$IllustrisTNG$\quad$ & \textbf{TNG-300} & \textbf{300} & $\mathbf{\quad 2 \times 2500^3 \quad}$ & $\mathbf{\quad 5.9 \times 10^7 \quad}$ & $\mathbf{\quad 1.1 \times 10^7 \quad}$ & \textbf{185} & $\quad$Stellar feedback, quasar-mode (thermal) and  \\
        
        & \textbf{TNG-100} & \textbf{100} & $\quad \mathbf{2 \times 1820^3} \quad$ & $\quad \mathbf{7.5 \times 10^6} \quad$ & $\quad \mathbf{1.4 \times 10^6} \quad$ & \textbf{92.5} & $\quad$radio-mode (kinetic) AGN feedback \\
        
        & \textbf{TNG-50} & \textbf{50}& $\mathbf{\quad 2 \times 2160^3 \quad}$ & $\mathbf{\quad 4.5 \times 10^5 \quad}$ & $\quad \mathbf{8.5 \times 10^4 \quad}$ & \textbf{36.3} & \\
       
         & \cellcolor{gray!20} TNG-300-2 &\cellcolor{gray!20}  300 & \cellcolor{gray!20} $\quad 2 \times 1250^3 \quad$ & \cellcolor{gray!20} $\quad 4.7\times 10^8 \quad$ & \cellcolor{gray!20} $\quad 8.8 \times 10^7 \quad$ & \cellcolor{gray!20} 375 & \cellcolor{gray!20} \\

         & \cellcolor{gray!20} TNG-100-2 & \cellcolor{gray!20} 100 & \cellcolor{gray!20} $\quad 2 \times 910^3 \quad$ & \cellcolor{gray!20} $\quad 6.0 \times 10^7 \quad$ & \cellcolor{gray!20} $\quad 1.1 \times 10^7 \quad$ & \cellcolor{gray!20} 185 & \cellcolor{gray!20} \\

          & \cellcolor{gray!20} TNG-50-2 & \cellcolor{gray!20} 50 & \cellcolor{gray!20} $\quad 2 \times 1080^3 \quad$ & \cellcolor{gray!20} $\quad 3.6 \times 10^6 \quad$ & \cellcolor{gray!20} $\quad 6.8 \times 10^5 \quad$ & \cellcolor{gray!20} 35 & \cellcolor{gray!20} \\
         
         & \cellcolor{gray!20} TNG-300-3 & \cellcolor{gray!20} 300 & \cellcolor{gray!20} $\quad 2 \times 625^3 \quad$ & \cellcolor{gray!20} $\quad 3.8 \times 10^9 \quad$ &  \cellcolor{gray!20} $\quad 7.0 \times 10^8 \quad$ & \cellcolor{gray!20} 757 & \cellcolor{gray!20} \\
         
         & \cellcolor{gray!20} TNG-100-3 & \cellcolor{gray!20} 100 & \cellcolor{gray!20} $\quad 2 \times 455^3 \quad$ & \cellcolor{gray!20} $\quad 4.8 \times 10^8 \quad$ & \cellcolor{gray!20} $\quad 9.0 \times 10^7 \quad$ & \cellcolor{gray!20} 370 & \cellcolor{gray!20} \\
         
         & \cellcolor{gray!20} TNG-50-3 & \cellcolor{gray!20} 50 & \cellcolor{gray!20} $\quad 2 \times 540^3 \quad$ & \cellcolor{gray!20} $\quad 2.9 \times 10^7 \quad$ & \cellcolor{gray!20} $\quad 5.4 \times 10^6 \quad$ & \cellcolor{gray!20} 145 &  \cellcolor{gray!20} \\
         
        \midrule
        
        Simba & \textbf{Simba-100} & \textbf{147} & $\mathbf{\quad 2 \times 1024^3 \quad}$ & $\mathbf{\quad 9.6 \times 10^7 \quad}$ & $\mathbf{\quad 1.82 \times 10^7 \quad}$ & $\mathbf{735}$ & $\quad$Stellar feedback, AGN-wind, AGN-jet, X-ray \\
        
        & Simba-50 & 74 & $\quad 2 \times 512^3 \quad$ & $\quad 9.6 \times 10^7\quad$ & $\quad 1.82 \times 10^7 \quad$ & 735 & $\quad$Stellar feedback, AGN-wind, AGN-jet, X-ray \\
        
        & No-X-ray & 74 & $\quad 2 \times 512^3 \quad$ & $\quad 9.6 \times 10^7 \quad$ & $\quad 1.82 \times 10^7 \quad$ & 735 & $\quad$Stellar feedback, AGN-wind, AGN-jet \\
        
        & No-jet & 74 & $\quad 2 \times 512^3 \quad$ & $\quad 9.6 \times 10^7 \quad$ & $\quad 1.82 \times 10^7 \quad$ & 735 & $\quad$Stellar feedback, AGN-wind \\
        
        & No-AGN & 74 & $\quad 2 \times 512^3 \quad$ & $\quad 9.6 \times 10^7 \quad$ & $\quad 1.82 \times 10^7 \quad$ & 735 & $\quad$Stellar feedback \\
        
        & No-feedback & 74 & $\quad 2 \times 512^3 \quad$ & $\quad 9.6 \times 10^7 \quad$ & $\quad 1.82 \times 10^7 \quad$ & 735 & $\quad$None \\
        
        & \cellcolor{gray!20} Simba-25 & \cellcolor{gray!20} 37 & \cellcolor{gray!20} $\quad 2 \times 256^3 \quad$ & \cellcolor{gray!20} $\quad 9.6 \times 10^7 \quad$ &  \cellcolor{gray!20} $\quad 1.82 \times 10^7 \quad$ & \cellcolor{gray!20} 735 & \cellcolor{gray!20} $\quad$Stellar feedback, AGN-wind, AGN-jet, X-ray \\
        
        & \cellcolor{gray!20} \cellcolor{gray!20} $\quad$Simba-HighRes$\quad$ & \cellcolor{gray!20} 37 & \cellcolor{gray!20} $\quad 2 \times 512^3 \quad$ &\cellcolor{gray!20}  $\quad 1.2 \times 10^7 \quad$ & \cellcolor{gray!20} $\quad 2.28 \times 10^6 \quad$ & \cellcolor{gray!20} 368 & \cellcolor{gray!20} $\quad$Stellar feedback, AGN-wind, AGN-jet, X-ray \\
        
        \bottomrule
        
        & & & & & & &\\
        
    \end{tabularx}
    {\footnotesize From left to right, the columns show: the parent project of the simulations considered (EAGLE, IllustrisTNG, Simba); the name of the individual runs within each project that have been adopted in this work;  the length of each side of the simulation box; the number of initial resolution elements, equally split between DM and gas; the mass resolution for the DM particles; the average mass of the initial gas elements; the late-time gravitational softening length for gas elements (see main text for details); a short description of the feedback modes active in each simulation (these vary across the Simba runs, whereas they are the fixed within the EAGLE and IllustrisTNG projects). The `flagship', or `fiducial', runs are indicated in boldface. The runs highlighted with a grey background have been utilised exclusively for testing the robustness of the physical interpretation of our results (see Appendix~\ref{app:robustness}), and not for the main analysis presented in this paper.}
\end{table*}

In this section, we describe the simulations that we adopted in this work. We will proceed in order of complexity with respect to the AGN feedback implementation, beginning with the EAGLE simulation (\S~\ref{sec:EAGLE}), which includes a thermal AGN feedback mode only. In \S~\ref{sec:TNG}, we will then describe the IllustrisTNG simulation, encompassing a dual thermal and kinetic AGN feedback model. Finally, in \S~\ref{sec:simba}, we will illustrate the different variants of the Simba simulation, containing up to three distinct AGN feedback channels. Since all simulations have been extensively described in the literature, we will provide only a brief summary of their characteristics, emphasising the aspects that are most relevant to our work, while referring the interested reader to the original papers for further details. The main specifics of the runs adopted in this work are listed in Table~\ref{tab:simulations}. In all cases, we consider the snapshots and halo catalogues at $z=4$, $z=2$, $z=1$, and $z=0$.

\subsection{EAGLE}
\label{sec:EAGLE}

EAGLE is a suite of cosmological hydrodynamical simulations run with a modified version of the \textsc{Gadget}-3 hydrodynamic code \citep{Springel_2005_bh, Springel_2008} known as \textsc{Anarchy} \citep{Schaller_2015_hydro}. The simulations start at redshift $z=127$, assuming a \LCDM\ cosmology consistent with the Planck-2014 observations \citep{Planck_2014}: $\Omega _{\rm{m}} = 0.307$, $\Omega _{\rm{\Lambda}} = 0.693$, $\Omega _{\rm{b}} = 0.04825$, $h = 0.6777$, $\sigma _{\rm{8}} = 0.8288$, and $n_{\rm{s}} = 0.9611$, with the usual definitions of the parameters. Initially, there is an equal number of DM and gas resolution elements, evolved as self-gravitating particles and via smoothed particle hydrodynamics (SPH), respectively. The gravitational softening length is fixed in comoving units up to $z=2.8$, and in proper units thereafter. The latter is quoted in Table~\ref{tab:simulations}.

EAGLE employs sub-grid modelling to simulate the physical processes that occur below the resolution limit, including radiative cooling and photoheating \citep{Wiersma_2009a}. Hydrogen reionization is driven by a spatially uniform and time-independent ultraviolet background (UVB) following \cite{HM01}. Star formation follows a pressure-based Kennicutt-Schmidt law with observationally fixed parameters, a metallicity-dependent density threshold \citep[see also][]{Schaye_2004, Schaye_2010}, and a temperature floor to mimic unresolved interstellar medium (ISM) physics \citep{DallaVecchia_2012}. Type Ia supernovae follow an exponential delay-time model consistent with observed rates \citep{Graur_2014} and eject metals based on W7 yields \citep{Thielemann_2003}. Energy injection from supernovae informs the broader stellar feedback scheme, which is modelled with a stochastic thermal approach \citep{DallaVecchia_2012} that overcomes overcooling by heating gas by a fixed temperature increase ($\Delta T = 10^{7.5} \K$) and modulating the injected energy fraction.

Black hole (BH) particles are seeded with an initial mass of \(1.48 \times 10^{5}\,\mathrm{M}_{\odot}\) at the centres of DM haloes with total masses exceeding \(1.48 \times 10^{10}\,\mathrm{M}_{\odot}\), provided that no BH is already present, following the method described by \cite{Springel_2005_bh}. The growth of the BH mass is governed by a modified Bondi-Hoyle accretion formalism that incorporates the relative motion of the surrounding gas \citep{Rosas_2015}, with the resulting accretion rate constrained by the Eddington limit. As accretion proceeds, feedback energy is accumulated in a reservoir and later deposited into nearby gas particles. This is done through a stochastic, thermally coupled AGN feedback model \citep{Booth_2009} with a temperature jump of $\Delta T_{\rm AGN} \geq 10^{8.5} \K$ and a calibrated coupling efficiency of $15\%$. 

Haloes are identified through a Friends-of-Friends (FoF; \citealt{Davis_1985}) algorithm with linking length equal to 0.2 times the mean inter-particle separation. Galaxies (substructures) are then defined by applying the \texttt{SUBFIND} algorithm \citep{Springel_2001_subfind, Dolag_2009} and removing gravitationally unbound particles. Halo and galaxy properties are saved in publicly available catalogues \citep{McAlpine_2016}, which we queried throughout this work. 

\subsection{IllustrisTNG}
\label{sec:TNG}

In the IllustrisTNG suite of simulations, DM is implemented in the form of self-gravitating Lagrangian particles, whereas gas resolution elements evolve an unstructured Voronoi tessellation following a fully magneto-hydrodynamic scheme based on the \texttt{Arepo} moving-mesh code \citep{Arepo}. The initial conditions are generated via the \texttt{N-GENIC} code \citep{Springel_2005} at $z=127$, assuming the \LCDM\ Planck-2016 cosmology \citep{Planck_2016} ($\Omega_{0}=0.3089$, $\Omega_{\rm b}=0.0486$, $\Omega_{\Lambda}=0.6911$, $\sigma_8 = 0.8159$, $n_{\rm s} = 0.9667$, and $h=0.6774$). Different softening lengths are adopted for the collisionless and gaseous components. For the latter, the softening adapts in size, scaling with the effective cell radius. Since in this work we are mostly concerned with the gas content of haloes, we report the minimum gas softening lengths in Table~\ref{tab:simulations}.

The galaxy formation framework incorporates both primordial and metal-line cooling mechanisms, employs sub-grid prescriptions for star formation, the ISM thermal state, and metal enrichment enrichment from AGB stars as well as Type Ia and Type II supernovae \citep{Vogelsberger_2013, Springel_2003}. The model for stellar feedback introduces isotropic wind injection, a redshift-dependent velocity with a minimum floor to regulate mass loading \citep{Torrey_2014}, and metallicity-dependent wind energy (including a thermal fraction) that strengthens feedback in low-mass galaxies. 

Supermassive BHs are seeded with an initial mass of $8\times10^5\,h^{-1}\,\mathrm{M}_\odot$ at the centres of haloes once they exceed a mass threshold of $ 5\times10^{10}\,h^{-1}\,\mathrm{M}_\odot$. Their subsequent growth follows the Bondi-Hoyle-Lyttleton prescription \citep{Hoyle_1939, Bondi_1944, Bondi_1952} as long as it exceeds a black-hole-mass dependent fraction of the Eddington accretion rate, which acts as a floor for the BH accretion rate. The accretion state of the BH determines which AGN feedback mode may be activated. During the high-accretion `quasar-mode', feedback energy is injected as thermal energy into the surrounding gas, representing the radiatively efficient coupling of accretion power \citep[e.g.][]{Springel_2005_bh, Vogelsberger_2013}. The thermal feedback energy rate is determined by the BH accretion rate and coupling efficiency and the resulting temperature increase depends self-consistently on the amount of energy injected and the local gas mass and density. In the low-accretion `radio-mode', feedback is delivered as kinetic energy via momentum injection along randomised directions, mimicking mechanically dominated outflows such as jets and winds in radiatively inefficient accretion regimes \citep[e.g.][]{Dunn_2006, Yuan_2014}. The injected kinetic energy is similarly tied to the BH accretion rate, but no fixed outflow velocity is imposed; instead, the resulting gas velocities follow from the available feedback energy and local gas conditions. This dual-mode approach captures the transition from thermally driven quasar feedback at high accretion rates to kinetically dominated radio feedback in massive haloes at low accretion rates.

Haloes are identified on the fly via a FoF algorithm with linking length equal to 0.2 times the inter-particle separation, while galaxies are extracted with the  \texttt{SUBFIND} algorithm \citep{Springel_2005, Dolag_2009}. We used the publicly available snapshots, and halo and subhalo catalogues made publicly available by \cite{TNGPublicDataRelease}.

\subsection{Simba}
\label{sec:simba}

Simba is a suite of cosmological simulations \citep{Simba} run with the \texttt{Gizmo} hydrodynamic code, in its meshless finite mass version \citep{Gizmo}. The gravitational softening length for all particle is fixed in comoving units. In Table~\ref{tab:simulations}, we report the maximum softening length ($z=0$). The simulations adopt a \LCDM\ cosmological model with parameters consistent with the Planck-2016 results : $\Omega_{\mathrm{m}}=0.3$, $\Omega_{\Lambda}=1-\Omega_{\mathrm{m}}=0.7$, $\Omega_{\mathrm{b}}=0.048$, $h=0.68$, $\sigma_8=0.82$, $n_s=0.97$. 

Simba uses the \texttt{Grackle-3.1} library \citep{Smith_2017} to model radiative cooling and photoionisation heating, accounting for metal-line cooling and non-equilibrium chemistry of primordial elements. The ionising UV background follows \cite{HM12} with self-shielding corrections from \cite{Rahmati_2013}. Star formation is modelled with the \cite{Schmidt_1959} $\rm H_2$-based relation following \cite{Krumholz_2011}, limited to dense molecular gas satisfying specific density and temperature criteria. Stellar feedback is implemented as star formation-driven galactic winds launched via kinetic, decoupled ejection, representing the combined effects of Type II supernovae, radiation pressure, and stellar winds. Outflows are generated perpendicular to the local velocity and acceleration, with mass loading and velocity calibrated to the scaling relations of the FIRE simulations \citep{Muratov_2015, Angles-Alcazar_2017b}. Winds are metal-enriched from supernova yields, with 30\% thermally heated and the rest ejected cold ($T\sim10^3$ K). Wind particles are hydrodynamically decoupled and cooling is disabled to avoid numerical artefacts, recoupling once they have propagated for sufficiently long time, reached low enough densities, or matched the surrounding gas properties.

Simba models BHs as sink particles that accrete gas through a dual scheme: hot, non-ISM gas is accreted via the Bondi prescription \citep{Bondi_1952}, while cooler gas within the accretion kernel follows a torque-limited model driven by gravitational instabilities \citep{Hopkins_2011, Angles-Alcazar_2013, Angles-Alcazar_2015, Angles-Alcazar_2017a}. AGN feedback operates in three modes depending on black hole mass and accretion rate. BHs with accretion rate above $0.2$ in Eddington units launch bipolar outflows aligned with the angular momentum of the gas within the BH kernel. Such `AGN-wind' mode provides kinetic feedback without direct thermal coupling. As the BH accretion rate drops below the aforementioned threshold, and the BH mass exceeds $10^{7.5} \Msun$, the `AGN-jet' mode is additionally activated. This mode produces faster, collimated outflows with increasing speed as the accretion rate declines. The velocity increase of ejected gas due to the AGN-jet mode is capped at $7000\ {\rm km \, s^{-1}}$ once the BH accretion rate falls below $0.02$ in Eddington units. Because jets are expected to carry highly energetic plasma, jet outflows are heated to the virial temperature of the halo \citep{Voit_2005}, consistent with observations of synchrotron-dominated jets thermalising hot gas near this temperature scale \citep{Fabian_2012}. Finally, when the host galaxy of BHs that are already undergoing AGN-jet feedback is gas-poor (gas fraction $<0.2$), the `X-ray' mode becomes active, depositing energy into gas within the BH kernel with a strength that decreases with distance from the black hole. This heating raises the temperature of diffuse, non-ISM gas and splits its effect on ISM gas between thermal input and kinetic acceleration, driving outward flows. As a result, X-ray feedback prevents overcooling of dense gas that could otherwise occur under the ISM pressurisation model.

On top of runs with the full Simba galaxy formation model, we also consider variants where at least one of the feedback modes above have been deactivated. These are all listed in Table~\ref{tab:simulations}. The full-physics runs Simba-100, Simba-50, Simba-25, and Simba-25-HighRes reflect the respective box sizes of $100 \hMpc$, $50 \hMpc$, and $25 \hMpc$, respectively. This is different from the naming convention adopted for all other simulations in Table~\ref{tab:simulations}, where the number reflects the box size without the $h^{-1}$ factor. Our choice to depart from such convention in the case of Simba is to be consistent with the labels adopted in previous literature \citep[e.g.][]{Simba, Sorini_2022, Sorini_2024}.

In all runs, halos are identified using a 3D FoF algorithm (linking embedded in 0.2 times the average inter-particle separation) \texttt{Gizmo}, derived from \texttt{Gadget-3}. Galaxies are coupled to haloes in post-processing, through the \texttt{yt}-based software \textsc{Caesar} \footnote{\url{https://caesar.readthedocs.io/en/latest/}}. The resulting catalogues containing halo and galaxy properties, as well as snapshot files, are publicly available and have been extensively utilised in this work.

\section{Model}
\label{sec:model}

\begin{figure*}
    \centering
    \includegraphics[width=\textwidth]{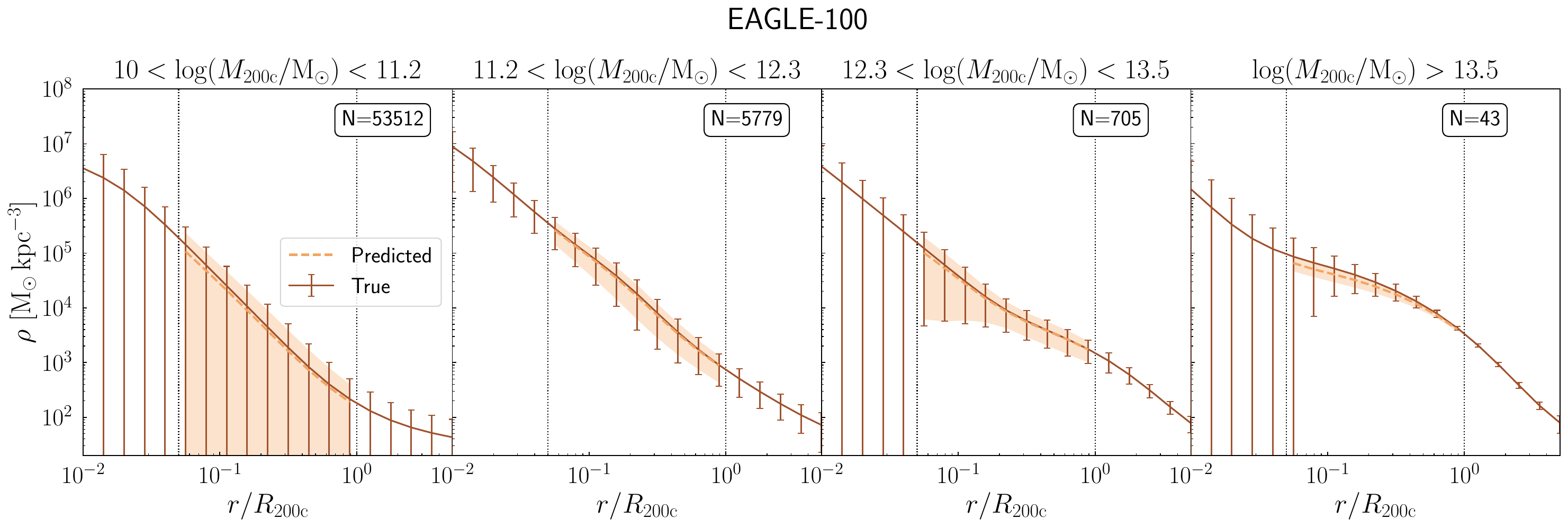}
    \includegraphics[width=\textwidth]{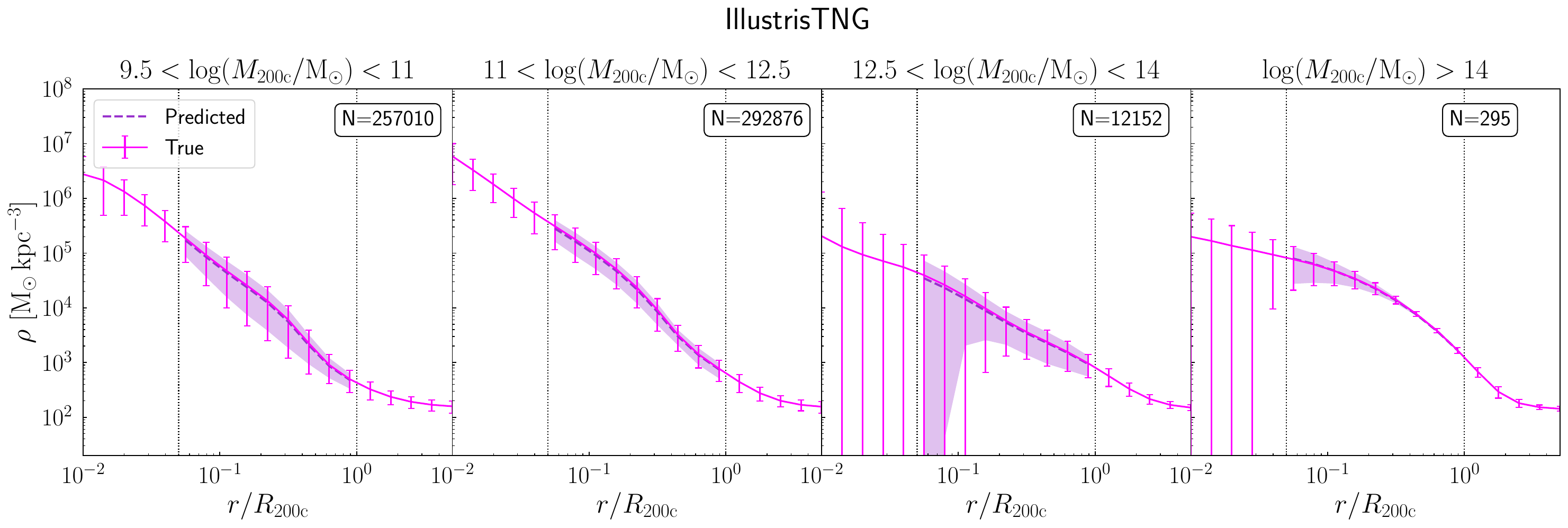}
    \includegraphics[width=\textwidth]{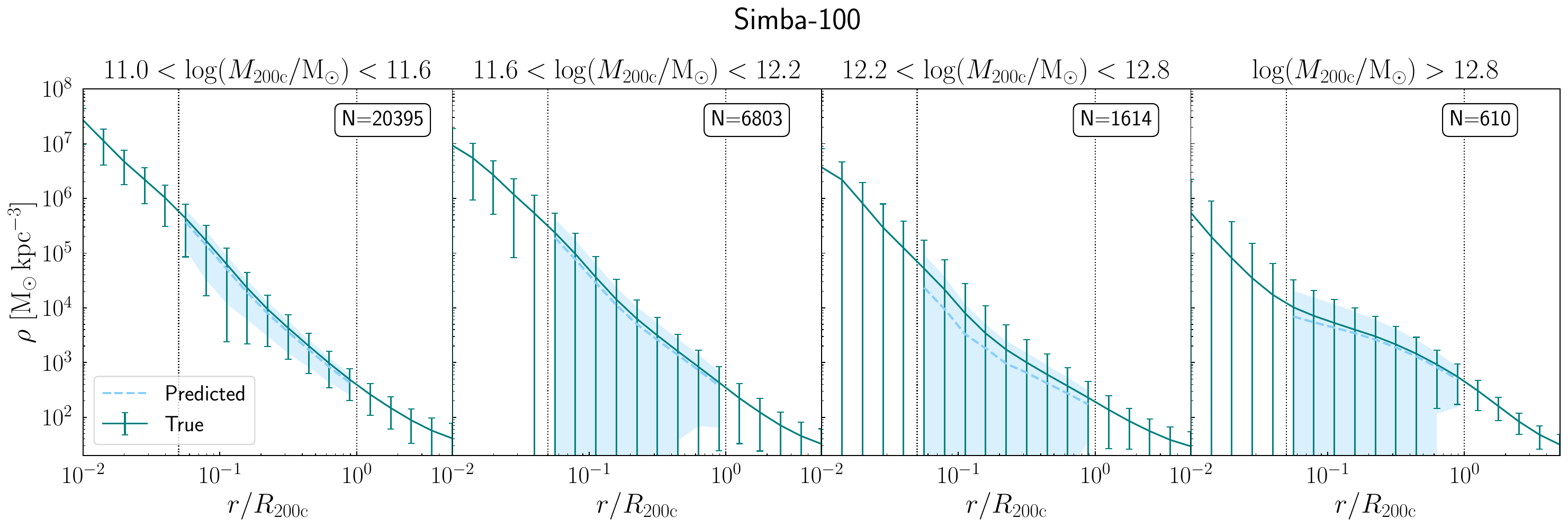}
    \caption{Spherically averaged radial profiles of the total gas density around haloes in the test sets extracted from the EAGLE-100, IllustrisTNG, and Simba-100 simulations (top, middle, and bottom panels, respectively) at redshift $z=0$. The solid lines show the mean gas density profile of haloes in the $M_{\rm 200c}$ bin indicated on top of every panel. The corresponding number of haloes is reported in the upper-right corner. The error bars represent the $16^{\rm th}-84^{\rm th}$ percentile spread across the halo sample. The dotted vertical lines mark the $0.05 \, R_{\rm 200c} - R_{\rm 200c}$ radial range, wherein we aim to recover the simulated density profiles through our RF algorithm. The mean of the predicted density profiles is shown by the dashed lines, and the shaded areas represent the $16^{\rm th}-84^{\rm th}$ spread in the predictions. In all flagship simulations considered, the RF algorithm accurately recovers both the mean and scatter of the gas density profiles around haloes over several orders of magnitude in $M_{\rm 200c}$. It performs less accurately in Simba-100 at the high-mass end.
    }
    \label{fig:dens_prof}
\end{figure*}

In this section, we construct a model to predict the halo gas density profile from the knowledge of its total mass and a set of properties of its central galaxy, under the feedback prescriptions. Conceptually, this task will be accomplished by a function, to be determined. Given the complex interplay among different galaxy and halo properties with their gaseous environment, such function will not be defined analytically from first principles. Instead, we will define it through a data-driven approach, based on training a ML algorithm on the density profiles and galaxy/halo properties data in each of the simulation presented in \S~\ref{sec:simulations}. We describe how we extract the relevant datasets in \S~\ref{sec:profiles}-\S~\ref{sec:features} and then explain the ML algorithm in \S~\ref{sec:algorithm}.

\subsection{Gas density profiles}
\label{sec:profiles}

We begin with describing how we constructed the radial profile of the total gas mass density in the haloes of the simulations considered. 

For a given snapshot of a given run, we first consider all main haloes in the corresponding catalogue (see \S~\ref{sec:simulations} for details). The catalogues report the position of the minimum of the gravitational potential for each halo. We will refer to this quantity as `halo centre' hereafter. The catalogues also provide the values of $R_{\rm 200c}$, defined as the halo-centric distance enclosing a total mass density equal to 200 times the critical density of the Universe. In the remainder of the paper, we shall refer to this scale length as $R_{\rm 200c}$ or ` virial radius' interchangeably and denote the enclosed total mass as $M_{\rm 200c}$.

For each halo in the catalogue, we organise the surrounding gas elements into 20 bins of radial distance from its centre. The first bin spans the radial interval between the halo centre and $0.01 \, R_{\rm 200c}$. The remaining 19 bins cover the range $0.01 - 5 \, R_{\rm 200c}$ with equal width in logarithmic space.\footnote{The exact number of bins is unimportant: we verified that with twice or half as many bins the accuracy of our ML algorithm (see \S~\ref{sec:validation}) vary by $\sim 1\%$ and the main conclusions of our work would be unchanged.} The gas density corresponding to each bin is then straightforwardly computed by dividing the total mass of the gas elements falling in that bin by the volume enclosed between the spherical shells defined by the boundaries of the bin. We stress that we consider \textit{all} gas elements in the simulations, not only those identified as bound to the halo by the FoF algorithm. 

To ensure the numerical robustness of the halo gas density profiles, we compute them for `well resolved' haloes only. The criterion that we adopt to deem a halo to be `well resolved' for our purposes is that it contains at least 5000 resolution elements as determined by the FoF algorithm. This rather stringent requirement, in line with choices made in previous related works \citep[e.g.][]{Schaller_2015, Sorini_2022, Sorini_2024}, ensures that there are enough gas resolution elements in each individual radial bin to cut down the Poisson noise in the corresponding gas density to an acceptable level. We further require that the central galaxy is not devoid of gas, so that we do not mistakenly select spurious dark subhaloes that might have been found by the subhalo-finding algorithm.

We show the gas density profiles obtained from the three flagship simulations in Figure~\ref{fig:dens_prof}, separated into four different $M_{\rm 200c}$ bins, as indicated above each panel. The halo mass bins differ from simulation to simulation, because the dynamic range of well resolved haloes varies depending on the mass resolution and box size. The mass range probed is particularly large in the case of IllustrisTNG. The reason is that, only for this simulation, we combine all haloes and density profiles selected from the TNG-50, TNG-100, and TNG-300 run in the same dataset. This enables us to take advantage of the higher resolution of TNG-50, hence probing lower-mass haloes, with the higher statistics of larger-mass haloes given by the larger boxes. Consequently, the IllustrisTNG mass bins generally contain more haloes than in their EAGLE and Simba counterparts, as annotated in the box within each panel. In the remainder of the paper, we will always combine the data of the IllustrisTNG flagship boxes together, hence we shall refer to them collectively as `IllustrisTNG'. This is the same approach adopted by \cite{Sorini_2025}, who showed that the concentration-mass relationships of haloes in overlapping mass ranges across the different TNG boxes are consistent with one another. We verified that this is the case for the gas density profiles as well.

For all simulations, we show the mean gas density in each radial bin, averaged over all haloes in a given $M_{\rm 200c}$ bin. The error bars show the $16^{\rm th}-84^{\rm th}$ percentile range of the halo-to-halo scatter of the gas density in each bin. By visual inspection, we can already identify qualitative differences among the profiles produced by the different simulations. Simba produces profiles resembling a power law, at least within the interval $0.01 \, R_{\rm 200c} - R_{\rm 200c}$, which, as we will explain later, will be the most relevant for our work. This result is consistent with the previous findings by \cite{Sorini_2024}. EAGLE gives rise to power-law profiles up to $M_{\rm 200c} \approx 10^{13.5} \, \rm M_{\odot}$, above which they appear to exhibit a core-like feature, potentially resembling a modified Navarro-Frenk-White profile \citep{mNFW_1, mNFW_2}. This is true for the IllustrisTNG simulation too, where this features persists at lower masses as well. We can see a transition between a shallower gas density profile within $\sim 0.1-0.2 R_{\rm 200c}$ and a steeper profile beyond this scale, becomes more evident in the mass range $10^{9.5} - 10^{12.5} \Msun$. 

It is not obvious to tell exactly what is the cause of the trends observed across different mass bins in the various simulations. Given that all simulations rely on very similar cosmological models, it is highly likely that the plurality of the predictions follows from the respective baryonic subgrid physics. Developing an intuition as to what galaxy properties shape the gas density profiles of haloes of different mass, at different redshifts, and under different feedback schemes, is exactly the goal of the algorithm that we intend to build. 

\subsection{Galaxy and halo properties}
\label{sec:features}

Having constructed the gas density profiles, we now need to select a subset of the properties of the respective haloes and their central galaxies to serve as the input variables for the ML model that we are going to build. In principle, we could select all properties recorded in the halo catalogues. However, an excessive amount of properties may lead to over-fitting. A more sensible approach is therefore to select a small number of properties that can still guarantee accurate predictions for the gas density profiles. Obviously, the accuracy of the algorithm can be assessed only a posteriori, and whether that is `good enough' depends on the intended application of the model. We will discuss this point in \S~\ref{sec:validation}; for now, all we can do is select properties that we expect to have a strong interplay with the halo gas density profiles, based on physical considerations. Because the gas distribution within haloes is highly influenced by feedback processes, we mostly turn our attention to galaxy properties that are directly connected to the implementation of feedback prescriptions across the simulations considered in this work. 

Here is the list of properties that we selected for our model, together with the rationale behind every choice.

\begin{itemize}
    \item The total gas mass of the central galaxy, $M_{\rm gas}$. The normalisation of the gas density profile certainly depends on the total gas mass within the halo, which we expect to be connected to $M_{\rm gas}$. In the case of the IllustrisTNG and EAGLE runs, $M_{\rm gas}$ is simply the total mass of all gas resolution elements belonging to the most massive substructure within the halo, as identified by the \texttt{SUBFIND} algorithm. This quantity is not present in Simba halo catalogues, which instead report the total mass of all gas elements within a spherical aperture of $30 \, \rm ckpc$ around the centre of the most massive substructure.

    \item The stellar mass, $M_{\star}$, and star formation rate (SFR), $\dot{M_{\star}}$, of the central galaxy. These quantities determine key characteristics of the galaxy, including how active it is (i.e., star-forming, quenched, or green-valley). In particular, the amount of stars going supernovae is proportional to the overall amount of stars, therefore we expect $M_{\star}$ and $\dot{M_{\star}}$ to bear some information on how stellar feedback shapes the gas distribution in the parent halo.

    \item The mass, $M_{\rm BH}$, and accretion rate, $\dot{M}_{\rm BH}$, of the BH in the bulge of the central galaxy. These properties are tied to the AGN feedback mechanisms in the simulations considered, for example by determining the amount of energy deposited in the surrounding gas (e.g., EAGLE thermal or IllustrisTNG quasar-mode AGN feedback) or the speed of outflows ejected in kinetic modes (e.g., IllustrisTNG radio-mode or Simba AGN-jet feedback). Thus, $M_{\rm BH}$ and $\dot{M}_{\rm BH}$ are poised to reflect the impact of AGN feedback on the halo gas density profile.

    \item The total halo mass, $M_{\rm 200c}$. This is the only property of the halo, rather than its central galaxy, that we consider. We choose it because different $M_{\rm 200c}$ scales correspond to physically different types of objects (i.e., host haloes of dwarfs, Milky-Way type galaxies, galaxy groups, clusters). Therefore, $M_{\rm 200c}$ should carry global information on the relevant physics impacting its own matter distribution, including the gas density profile.
    
\end{itemize}

We now have all necessary data to build our ML algorithm, which will be the subject of the next subsection.

\subsection{Algorithm}
\label{sec:algorithm}

Our aim is to develop a data-driven model that provides an estimate of the halo gas density profile at each radial bin, given the properties listed in \S~\ref{sec:features}. However, it is well established that the finite mass resolution of N-body simulations can lead to artificial features in the central density profiles of haloes. The profiles can be considered numerically robust outside the so-called `convergence radius'. Adopting the criterion proposed by \cite{Power_2003} and following the estimator introduced by \cite{Ludlow_2019}, we verified that the convergence radius typically falls at $\sim 5\%$ the virial radius in all simulations considered. Therefore, we will restrict our predictions of the gas density profiles to the radial bins within the range $0.05 < r/R_{\rm 200c} < 1$. 

The estimator will be given by a random forest (RF) algorithm, which needs to be trained on the halo and galaxy properties, and corresponding halo gas density profiles, extracted from each simulation. Adopting the typical ML terminology, these are the \textit{input features} and \textit{target variables} of our model, respectively. Since a RF is an ensemble of simpler algorithms called `decision trees', we will first describe these in \S~\ref{sec:decision_trees}. Then, we will illustrate how they are combined into RFs in \S~\ref{sec:RF}. Finally, we will explain how we train our algorithm in \S~\ref{sec:training}.

\subsubsection{Decision trees}
\label{sec:decision_trees}

Decision trees are widespread ML models that can be used either as classification or regression algorithms. Our aim is an example of regression, since both our input features and target variables can assume continuous rather than discrete values. Therefore, we hereby summarise the main aspects of decision trees in the context of regression only.

The starting point of a decision tree is to select one input feature, fix a threshold for such feature, and split all data in the training set (\textit{root node}) based on whether they lie above or below the threshold. Both the feature and the corresponding threshold are chosen such that the variance of the target variables associated with the two data subsets (\textit{nodes}) after the split is minimised. The procedure is then repeated for each node, and re-iterated until one of the following criteria is met: the variance of the target variables associated with the data in a given node falls below a pre-determined threshold; splitting that node would exceed a pre-fixed maximum number of splits (\textit{depth}); the node contains less than a minimum number of data elements, which is set in advance.

Once the algorithm stops, one can fix the values of the input features, and query the corresponding estimate of the target variables. The model then goes through all splits previously determined by the decision tree, from the root node down to the end nodes (\textit{leaves}). When a leaf is reached, the algorithm returns the mean of the target variables associated with the training data within that leaf. 

Decision trees are generally accurate and efficient ML algorithms. However, the low variance in their predictions comes at the cost of a high bias, meaning that they may not be robust to generalisation. For example, querying a decision tree with input feature values that are significantly different from those in the original dataset may lead to inaccurate estimates of the target variables. This issue can be addressed by considering a properly randomised ensemble of decision trees, of which RFs are an example.

\subsubsection{Random forests}
\label{sec:RF}

A RF algorithm requires running multiple decision trees. Each tree is trained on a different variant of the training set, constructed by re-sampling with replacement. At every node in each tree, the dataset split is determined by considering only a random subset of all available input features. The number of selected features is typically set to be approximately equal to the square root of the total number of features. Adhering to this convention, we select a subset of 2 out of the 6 input features listed in \S~\ref{sec:features}.

After the training is completed, the estimated target variables associated with a set of input features is determined by averaging over all estimates obtained by querying each tree in the ensemble. Clearly, this will result in a larger variance with respect to the results obtained with a single decision tree. But the RF is more robust to generalisation, meaning that the variance of its predictions will be unlikely to increase significantly when further examples are added in the training set.

RFs have been already applied on cosmological hydrodynamical simulations and proved to be accurate and efficient algorithms for studying the physics of the CGM \citep[e.g.][]{Appleby_2023_ML}. Though, as most ML algorithms, their accuracy can depend quite heavily on the dynamic range covered by the training set. We will discuss how we prepare our datasets for the training of the models next.

\subsubsection{Algorithm training}
\label{sec:training}

Before training the RF algorithm, we apply a practice known as \textit{feature engineering} to our dataset. This procedure consists in properly modifying the features to improve the accuracy of the model. The motivation is that ML algorithms are known to underperform if the dynamic range of their input features or target variables span several orders of magnitude. This is certainly the case of the gas density density, which can vary up to $\sim 3.5 \, \rm dex$ in the radial range of interest ($0.05 \leq r/R_{\rm 200c} \leq 1$). Similar considerations apply to the galaxy and halo properties listed in \S~\ref{sec:features}.

We therefore replace the values of $M_{\rm gas}$, $M_{\star}$, and $M_{\rm 200c}$ in our datasets with their decimal logarithm. However, the SFR, the BH mass and BH accretion rate are equal to zero for some elements in the training set, making a logarithmic rescaling unsuitable. Instead, we operate the following replacement:
\begin{equation}
\label{eq:archsinh}
    \tilde{X} \longmapsto  \log(e)\,  \mathrm{arcsinh} \left(\frac{\tilde{X}}{2\tilde{X}_0} \right) \, ,
\end{equation}
where $\tilde{X}$ represents any of the aforementioned potentially problematic features, $\log$ is the decimal logarithm, and $\rm archsinh$ the inverse hyperbolic sine function. The constant $\tilde{X}_0$ serves an appropriate re-scaling, to make the argument of the function dimensionless. It is set to $\tilde{X}_0 = 1 \, \rm M_{\odot}$ if $\tilde{X}$ has the dimensions of mass and $\tilde{X}_0 = 1 \, \rm M_{\odot} \, yr^{-1}$ if $\tilde{X}$ represents an accretion rate. 

\begin{figure}
    \centering
    \includegraphics[width=\columnwidth]{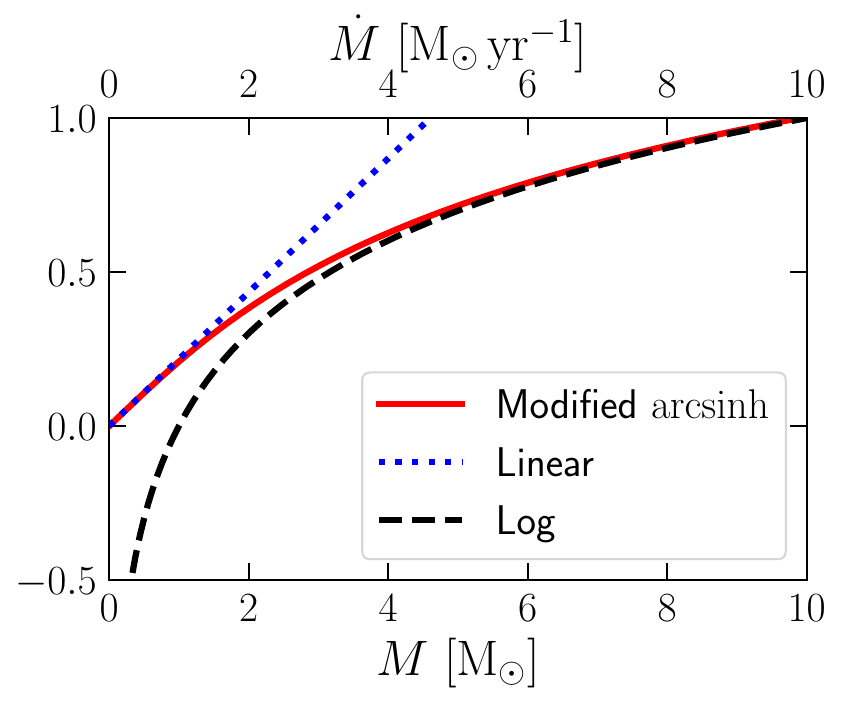}
    \caption{Modified inverse hyperbolic sine function (solid line) as defined in equation~\eqref{eq:archsinh}, which we applied to a subset of the input features ($\dot{M}_{\star}$, $\dot{M}_{\rm BH}$, $M_{\rm BH}$) to improve the performance of the RF algorithm (see \S~\ref{sec:features} for details). For masses above $\sim 7 \Msun$ (or accretion rates above $\sim 7 \Msun \rm \, yr^{-1}$), the function asymptotes to the decimal logarithm. As the property in question approaches zero, the function is well approximated by a linear relationship. In this figure, we zoom in the region where the smooth transition from linear to logarithmic regimes occurs. Crucially for properties such as the SFR, BH mass and accretion rate, the function is well defined at zero, thus making it preferable over a pure logarithmic function. At the same time, the scaling will be effectively logarithmic for large values of the input features. For example, the black hole mass can assume values ranging from zero to as large as $\sim 10^{10} \Msun$; the modified inverse hyperbolic sine is well defined in zero, while effectively scaling the BH mass logarithmically for $M_{\rm BH} \gg 7 \, \Msun$.
    }
    \label{fig:arcsinh}
\end{figure}

The advantage of the function in equation~\eqref{eq:archsinh} is that it guarantees a smooth transition from a linear transformation for $\tilde{X}/\tilde{X}_0 \ll 1$, to a logarithmic one for $\tilde{X}/\tilde{X}_0 \gg 1$. Hence, the rescaling in question is well defined for both null and positive values on the input feature. This can be seen in Figure~\ref{fig:arcsinh} and verified explicitly by recalling the definition of the inverse hyperbolic sine function,
\begin{equation}
    \mathrm{arcsinh} (x) = \ln (x  + \sqrt{x^2  + 1}), \,
\end{equation}
and taking the appropriate limits of equation~\eqref{eq:archsinh}:
\begin{align}
    \tilde{X} &\rightarrow \frac{\log(e)}{2} \frac{\tilde{X}}{\tilde{X}_0} \quad \quad \mathrm{for} \quad \frac{\tilde{X}}{\tilde{X}_0} \ll 1 \\
    \tilde{X} &\rightarrow \log \frac{\tilde{X}}{\tilde{X}_0} \quad \quad  \quad  \;\, \mathrm{for} \quad \frac{\tilde{X}}{\tilde{X}_0} \gg 1 \, .
\end{align}
In practice, for $\tilde{X}/\tilde{X}_0 \gtrsim 7$, the modified inverse hyperbolic sine function introduced in equation~\eqref{eq:archsinh} already matches the decimal logarithm within 1\%. This means that the BH mass, which can be as small as zero and as large as $\sim 10^{10} \Msun$ in the simulations considered, is effectively scaled logarithmically for $M_{\rm BH} \gg 7 \Msun$ while still be well defined in zero. Similarly, the SFR and BH accretion rate transition to a logarithmic scaling when they assume values $\gtrsim 7 \, \Msun \, \rm yr^{-1}$.

Although not strictly necessary, since all the values of the gas density profile are positive in our case, we still apply the same rescaling to the target variables, setting $\tilde{X}_0$ equal to the critical (comoving) density of the Universe. The reason is that in the future we plan to extend our model to predict not only the total gas density profile, but even the density profile of different gaseous phases (i.e., cold ISM, cool/warm/hot CGM, etc.) in different halo-centric radial bins. The density of one or more phases may be found to be equal to zero at certain distances in some haloes, making a logarithmic rescaling again unsuitable. Thus, by testing our model with the transformation in equation~\eqref{eq:archsinh} for both input features and target variables, we are confident that we can easily generalise it in future applications.

At this point, we can train our RF algorithm on the transformed input features and target variables, for every snapshot of each simulation. In all cases, we randomly select 80\% of the gas density profiles and corresponding halo and galaxy properties. This constitutes the training set; the remaining 20\% of the data are our test set. To tune the RF hyperparameters, the training set is further split into two subsets containing 80\% and 20\% of the data. The smaller subset serves as the validation set. The hyperparameters of our model are: the minimum number of data in each node that is required to trigger a split; the maximum depth of each decision tree in the ensemble; the number of decision trees. We varied these quantities in the ranges $2 - 64$, $2 - 64$, and $10-1000$, respectively. For each choice, we train the model on the larger subset of the training data and assess the accuracy (see \S~\ref{sec:validation} for the exact definition) on the validation set. We find that for all simulations considered, the best performance is achieved when the minimum data points in each node is set to $2$, the maximum depth fixed at $32$, and the number of estimators to $500$. The final model is then trained on the full training set using these selected hyperparameters.

\begin{figure*}
    \centering
    \includegraphics[width=\textwidth]{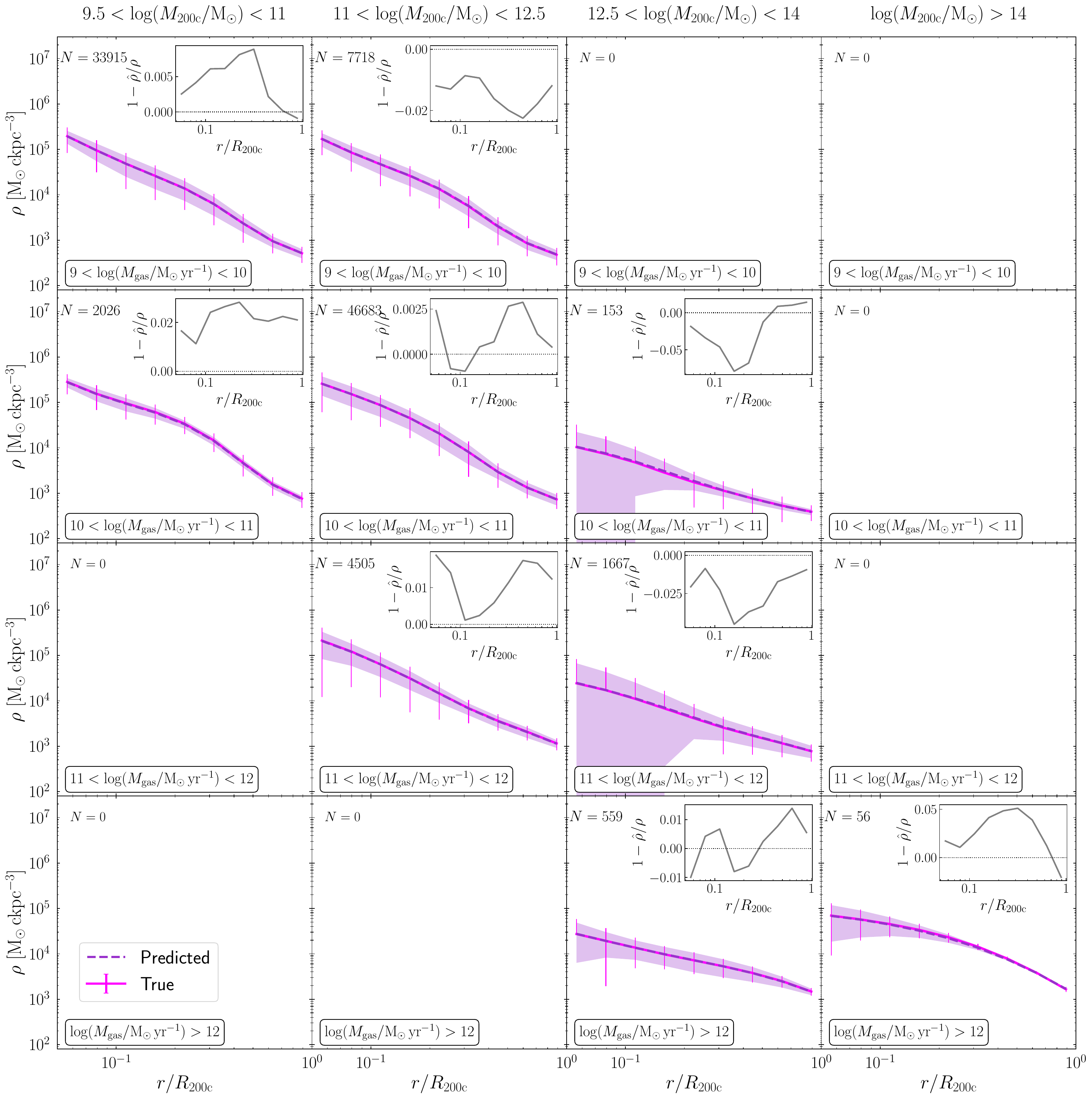}
    \caption{Spherically averaged comoving radial profiles of the total gas density around haloes in the test set extracted from IllustrisTNG simulations at redshift $z=0$. Each panel refers to the gas density profiles of haloes within the $M_{\rm 200c}$ bin indicated in the top of the column, hosting a central galaxy with total gas mass in the range annotated within the panel itself. The number of haloes falling within the 2D bin is reported in the upper-left corner of the panel. The profiles obtained directly from the simulations, and the predictions of the RF algorithm, following the same colour coding, line styles, and error bars as in the middle panel of Figure~\ref{fig:dens_prof}. The inset within every panel shows the geometric mean of the ratios between the predicted and true gas density profiles in the corresponding $M_{\rm 200c}$ -- $M_{\rm gas}$ bin, over the relevant radial range. Our RF algorithm generally recovers the true gas density profiles within 10\%. Considering the low statistics of input features, and the wide dynamic range of the gas density profiles, spanning up to four orders of magnitude, this is a remarkable achievement.}
    \label{fig:2D_dens_prof}
\end{figure*}

\section{Validation}
\label{sec:validation}

To assess the performance of our model, we need to compare the predicted gas density profiles of the haloes in the test set against the corresponding profiles obtained directly from the simulation. This needs to be done for all simulations and for every snapshot, so that we can evaluate the accuracy of the model as a function of redshift, and under different feedback scenarios.

We begin by visually inspecting the gas density profiles at $z=0$ for the EAGLE-100, IllustrisTNG, and Simba-100 simulations. The profiles obtained from these simulations were already shown with in Figure~\ref{fig:dens_prof}, where we split them into different $M_{\rm 200c}$ bins. The mean profiles predicted by our RF algorithm for the haloes in the test set, divided across the same mass bins, are shown with the dashed lines. The shaded areas indicate the $16^{\rm th}-84^{\rm th}$ percentile spread across the samples of predicted gas density profiles.

Notably, the predicted mean and simulated mean profiles appear to be generally in good agreement. Similarly, the scatter in the predictions is comparable with the $16^{\rm}-84^{\rm th}$ percentile error bars in the profiles taken from the simulations. This preliminary visual comparison suggests that our model can accurately reproduce \textit{both} the mean \textit{and} scatter of the gas density profiles. The only exception seems to be the Simba-100 simulations in haloes with $M_{\rm 200c} > 10^{12.2} \Msun$, where our model systematically underpredicts the mean gas density profile by $\sim 0.5 \, \rm dex$. We verified that this is the case only at $z=0$, whereas at higher redshift the agreement between model and simulations is consistently good over all halo mass bins. The drop in the accuracy at the higher-mass end may suggest that the input features are not sufficient to capture the full complexity of the Simba-100 feedback physics at low redshift (see discussion in \S~\ref{sec:discussion}). However, one should also bear in mind that the highest-mass bin contains different objects, ranging from Milky-Way host haloes to groups and clusters. The low statistics of clusters may be insufficient to properly train the RF algorithm at the high-mass end, hence lowering the overall accuracy in the highest-mass bin in Figure~\ref{fig:dens_prof}.

In fact, even narrow $M_{\rm 200c}$ bins would contain a variety of objects, such as gas-rich and gas-poor central galaxies alike, with active or quiescent AGN feedback. Thus, although Figure~\ref{fig:dens_prof} is reassuring in terms of the overall performance of our model, we should have a closer look at the accuracy of the predictions when we split the training set both with respect to $M_{\rm 200c}$ and some other input feature. This is what we do in Figure~\ref{fig:2D_dens_prof}, where, as an example, we show the simulated and predicted gas density profiles for haloes in the test set of the IllustrisTNG simulation at $z=0$. Each column of panels corresponds to the halo mass bin specified in the top part of the figure, while every row refers to the same $M_{\rm gas}$ bin, as annotated in each panel. The colour coding, line styles, shaded areas and error bars have the same meaning as in Figure~\ref{fig:dens_prof}. Additionally, to make the plots more easily legible, we omit the lower error bar whenever the $16^{\rm th}$ percentile of the simulated gas density profile at a given radial bin falls below the lower bound of the $y$-axis. We also restrict the plotted data to $0.05 \leq r/R_{\rm 200c} \leq 1$, which is the radial range of interest for our analysis. Within each panel, we add an inset reporting the relative difference between the geometric mean of the predicted gas density profile $\hat{\rho}$, and the same quantity given by the simulation, $\rho$. 

We find that in all $M_{\rm 200c}$-$M_{\rm gas}$ 2D-bins considered, the predicted mean gas density profile in the test set typically matches the true profiles within $\sim 5-10\%$. This level of accuracy is remarkable, considering the relative simplicity of the algorithm, the modest number of input features, and the wide dynamic ranged spanned by the gas density profiles (up to $3 \, \rm dex$ in the radial range considered). We verified that at higher redshift the mean accuracy improves even more. However, other simulations such as Simba-100 and its variants exhibit an overall lower accuracy, as we will discuss shortly. 

The dynamic range in the density profile suggests that the relative difference between predicted and true gas density may not be the most sensible proxy for the accuracy of our model. A more suitable metric would be the difference in the logarithm of the two quantities, because that would weigh the same relative difference on different scales equally. Also, it is fairer to assess the accuracy of our algorithm on its actual target variables, which are the logarithms of the density profiles\footnote{Technically, as explained in \S~\ref{sec:training}, the target variables would be the modified inverse hyperbolic sine as defined in equation~\eqref{eq:archsinh}. But as discussed in the same section, this function is very well approximated by a logarithm for all gas density profiles.}. Therefore, we now re-consider the same gas density profiles as in Figure~\ref{fig:2D_dens_prof}, divided amongst the same 2D bins, and calculate the accuracy of our model, as a function of radial distance. We now define the accuracy of the model as
\begin{equation}
    \label{eq:accuracy}
    \alpha(r) = 1- \left\vert \log \frac{\hat{\rho}(r)}{\rho (r)} \right\vert \, ,
\end{equation}
so that it equals unity if and only if the predicted and true gas density ($\hat{\rho}$ and $\rho$, respectively) at the distance $r$ are the same. We note that in principle, with the definition above, there is no lower limit to $\alpha(r)$, which could even become negative. However, as we will now show, in practice this never happens for our model.

\begin{figure*}
    \centering
    \includegraphics[width=\textwidth]{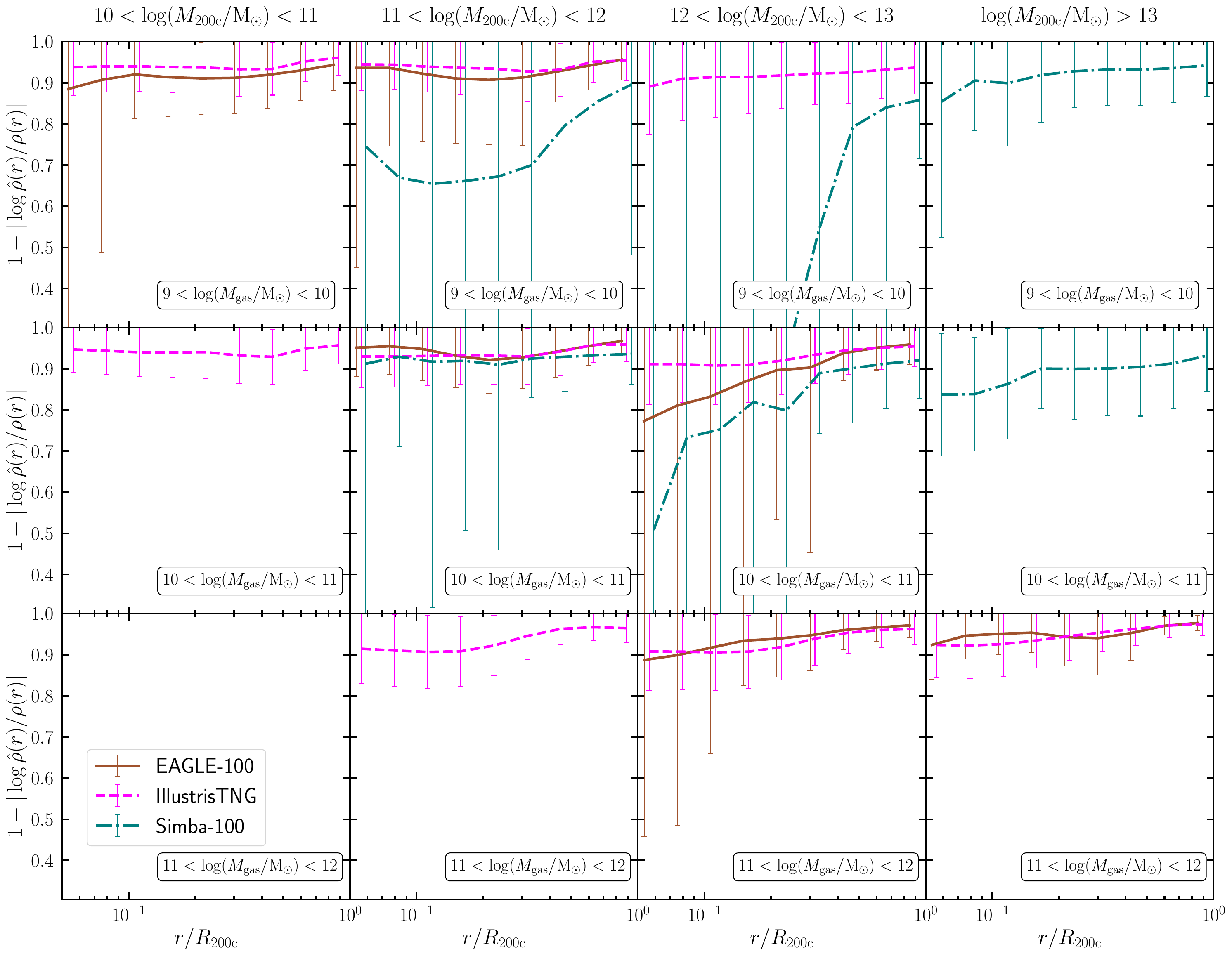}
    \caption{Average accuracy of the RF algorithm, as defined in \S~\ref{sec:validation}, at reproducing the gas density profiles in the test sets of the EAGLE-100, IllustrisTNG, and Simba-100 simulations at redshift $z=0$, as a function of the radial distance from the centre of the halo. The error bars indicate the $16^{\rm th}-84^{\rm th}$ percentile spread of the accuracy within each 2D-bin of $M_{\rm 200c}$ and $M_{\rm gas}$, the bounds of which are indicated in the upper part of the figure and in the annotation within each panel, respectively. The highest mean accuracy is above 90\% for EAGLE-100 and IllustrisTNG, and about 85\% for Simba-100. In general, the accuracy does not vary significantly over radial distance, although it can decay closer to the halo centre in specific 2D bins, especially in the case of Simba-100.
    }
    \label{fig:2D_accuracy}
\end{figure*}

Each panel in Figure~\ref{fig:2D_accuracy} then shows the accuracy introduced in equation~\eqref{eq:accuracy}. The thick lines represent the mean accuracy in each radial bin, averaged over all haloes in the $M_{\rm 200c}$-$M_{\rm gas}$ bin associated with each panel. The error bars indicate the $16^{\rm th} - 84^{\rm th}$ percentile spread in the accuracy at the radial bin considered. A small scatter on the position of the radial bins has been introduced to more easily distinguish the results relative to different simulations. However, the radial bins are the same across all runs considered.

The average accuracy for the IllustrisTNG simulations is consistently above $90\%$, with scatter limited within $\sim 10\%$. EAGLE-100 performs similarly well in all 2D-bins considered, while exhibiting lower accuracy ($\sim 78\%$) in the inner radial bins for haloes with $10^{10} < M_{\rm gas} / \mathrm{M}_{\odot} < 10^{11}$ and $10^{12} < M_{\rm 200c} / \mathrm{M}_{\odot} < 10^{13}$. Also, the scatter around the average accuracy tends to be bigger closer to the halo centre in most 2D-bins. Regarding Simba-100, the accuracy is above $65\%$ at all radial distances below $M_{\rm 200c}=10^{12} \Msun$ but can decay below 30\% in the innermost bins at higher masses. The accuracy raises back to $\sim 80\%$ in the highest halo mass bin. We will discuss possible causes for the sub-optimal accuracy in Simba-100 compared to the other simulations as well as future prospects for improvement in \S~\ref{sec:discussion}. 

From the definition in equation~\eqref{eq:accuracy}, the accuracy shown in Figure~\ref{fig:2D_accuracy} is the same whether the predicted gas density profile overestimates or underestimates the true profile by the same factor. As such, our accuracy estimator is blind to possible systematics in the predictions. Following what we have done for Figure~\ref{fig:2D_dens_prof}, we thus computed the relative difference of the geometric mean of the predicted and true gas density profiles within each 2D-bin in Figure~\ref{fig:2D_accuracy}. While within some of the 2D-bins our predictions systematically overestimate or underestimate the true profiles, we find that there is no systematic bias overall.

We showed the accuracy within 2D bins of halo mass and galaxy gas mass because these turned out to be the most important features in shaping the predictions of the gas density profile provided by our model (\S~\ref{sec:importance}). However, we verified that the results are qualitatively similar if we construct 2D-bins by pairing $M_{\rm 200c}$ with an input feature other than $M_{\rm gas}$. This means that, on average, our model does a good job of reproducing the gas density profiles at all radial distances, no matter which subsample of the training set we are considering. We also verified that at higher redshifts the accuracy improves for all simulations.  

\begin{figure}
    \centering
    \includegraphics[width=\columnwidth]{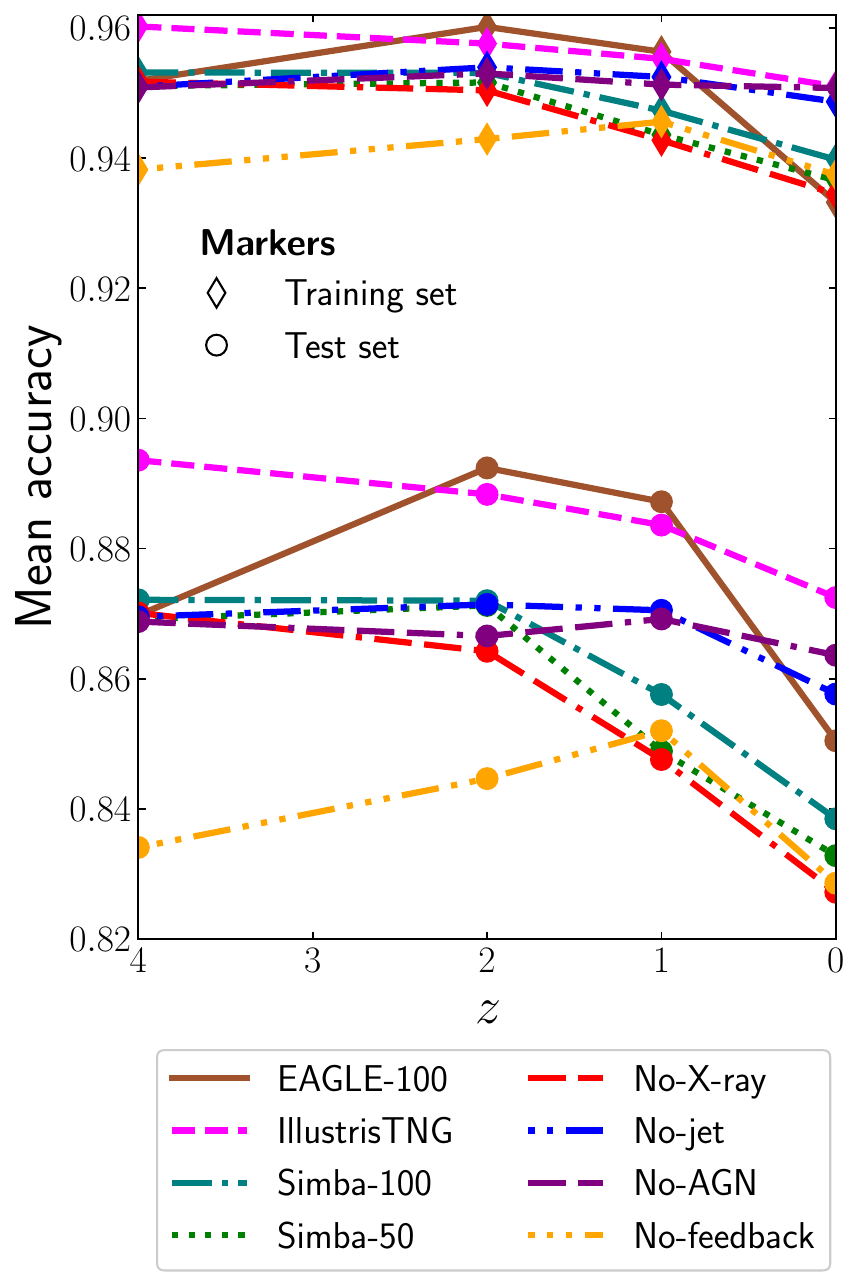}
    \caption{Overall accuracy of the RF algorithm at reproducing the gas density profiles in all simulations considered (each indicated with a different line style and colour), averaged over all radial bins and all haloes, as a function of redshift (see \S~\ref{sec:validation} for details). The diamonds show the results for the training set: the accuracy is above 92\% for all simulations, highlighting the self-consistency of the algorithm. The circles show the results for the test set: all flagship simulations maintain a high accuracy ($\sim$85\% for Simba-100, $\sim$90\% for EAGLE-100 and IllustrisTNG) down to $z=1$, with a few-percent decay at $z=0$. Nevertheless, this test proves that the algorithm can reproduce with high fidelity density profiles over which it was not trained, hence underscoring its predictive power.
    }
    \label{fig:accuracy_z}
\end{figure}

In this respect, it would be desirable to condense the accuracy of the model analysed in Figure~\ref{fig:2D_dens_prof} into one number that evaluates the overall performance of the algorithm for a given redshift and simulation. We can do this by averaging $\alpha(r)$ as defined in equation~\eqref{eq:accuracy} over all haloes in the simulation considered, and all all radial bins:
\begin{equation}
    \langle \alpha \rangle = 1 - \frac{1}{N_{\rm h}} \sum_{i=1}^{N_{\rm h}} \frac{1}{N_r} \sum_{j=1}^{N_r} \left\vert \log \left( \frac{\hat{\rho}_{ij}}{\rho_{ij}} \right) \right\vert \, ,
\end{equation}
where $\rho_{ij}$ and $\hat{\rho}_{ij}$ are, respectively, the true and predicted gas density at the $j$-th radial bin in the $i$-th halo; $N_{\rm h}$ and $N_r$ are the total number of haloes in the simulation considered, and the number of radial bins, respectively. We will refer to $\langle \alpha \rangle$ as `mean accuracy' hereafter.

We plot the mean accuracy as a function of redshift for all the flagship simulations, as well as all variants of Simba, in Figure~\ref{fig:accuracy_z}. Diamonds show the mean accuracy when computed by comparing the predicted and true gas density profiles in the training set. This is simply a self-consistency check, as any data-driven model is expected to accurately recover the target variables on which it was trained, starting from the same input features. Circles show the redshift evolution of the mean accuracy, as computed on the test set. This is the relevant parameter to assess the performance of the model, since it becomes practically useful when applied to data that the algorithm has not yet seen. For each simulation, trends are similar to those observed for the training set, but the overall accuracy is systematically lower. This is typical in ML models. 

The IllustrisTNG and EAGLE-100 simulations exhibit an overall better accuracy, close to $90\%$ for $z\gtrsim 1$, and around $86\%$ at $z=0$. In the case of Simba, the mean accuracy is lower: about $86-87\%$ at $z \gtrsim 2$ in all variants considered. This decays slightly at lower redshift, with a spread in accuracy between 83\% and 86\% depending on the run. The reduction in accuracy when switching from the training to the test set as the term of comparison for our predictions is around 5\%-7\% for IllustrisTNG and EAGLE, and 6-11\% for Simba.

The systematically lower accuracy for the Simba simulations suggests that the model would perhaps benefit from extra input features. We will discuss possible ameliorations in this sense in Section~\ref{sec:limitations}. More broadly, the accuracy declines at $z=0$ for all models, again hinting towards some important piece of information that the model is missing. However, we remind the reader that here we aim at presenting a proof of concept, which will be duly generalised in future work. It is still remarkable that with a simple model, based on a handful of input features, we can recover the true gas density profiles with $83-90\%$ accuracy in most cases.

This leads to the question of what level of accuracy can be considered `good enough'. Clearly, that depends on the specific application of the model that one has in mind, therefore there is no universal answer. An obvious application concerns analytical or semi-analytical models of galaxy formation where the star formation history in a galaxy is linked to the gas density profile of the host halo \citep[e.g.][]{HS03, Lacey_2016, SP21}. Our algorithm can be embedded in such frameworks to provide a simulations-informed connection between the properties of the central galaxy with the surrounding gas distribution. The desirable accuracy would then ultimately depend on the observations the specific star formation model is compared with. Therefore, the suitability of our algorithm as an ingredient of analytical or semi-analytical frameworks should be assessed on a case-by-case basis.

A more ambitious, forward-looking application would be to compare the predictions of our model with observations of the inferred density profile of several gaseous phases in different objects, from galaxies to clusters. As a reference, current observations of the X-ray surface brightness around low-redshift clusters provide estimates of the underlying hot gas density profile of $\sim 15\%$ \citep{Zhang_2024}. Although this does not necessarily translate into a 1-to-1 relationship with the total gas density profile studied in this work, $15\%$ appears to be a sensible benchmark to aim for. Our model performs better than that at redshifts $z \gtrsim 1$. At $z=0$, this is still the case for EAGLE-100 and IllustrisTNG, but not for the Simba runs. We can then conclude that over most of the redshift interval probed, and in the variety of feedback prescriptions considered in this work, our model seems promising for future applications on observational data. This adds a practical advantage to our model, on top of its theoretical significance.

\section{Sensitivity analysis}
\label{sec:sensitivity}

Data-driven models are occasionally subject to some degree of pushback from the scientific community, being regarded as `black boxes' with limited explanatory power \citep[e.g.][]{Rudin_2018}. However, there are statistical tools that enable us to easily interpret the predictions of theoretical models, including those based on ML approaches. In particular, we will exploit sensitivity analysis \citep{Fisher_1918, Sobol_1993, Saltelli_2010} to quantify the importance of each input feature on the predictions of our model, as well as the degeneracy among different features. 

Sensitivity analysis has already been successfully applied in astrophysics, for example to analyse the importance of the input parameters of the \texttt{GALFORM} \citep{Galform} semi-analytic model of galaxy formation \citep{Oleskiewicz_2020, Elliott_2021, Bose_2025}. Other works employed statistical analysis onto semi-analytic models or emulators, coupled with suites of N-body simulations \citep{Facundo_2012, Euclidemulator_2019, Nouri-Zonoz_2025}. To the best of our knowledge, our work constitutes the first application of Sobol sensitivity analysis to full cosmological hydrodynamical simulations.

In \S~\ref{sec:sensitivity} we provide the necessary mathematical background, in particular regarding the concept of Sobol indices \citep{Sobol_2001, Saltelli_2010}, which are instrumental for the main conclusions of our work. We will then describe our numerical implementation for computing Sobol indices in \S~\ref{sec:sampling}. The results of our sensitivity analysis will be discussed in the next section.

\subsection{Mathematical background}
\label{sec:sobol}

\begin{figure*}
    \centering
    \includegraphics[width=\textwidth]{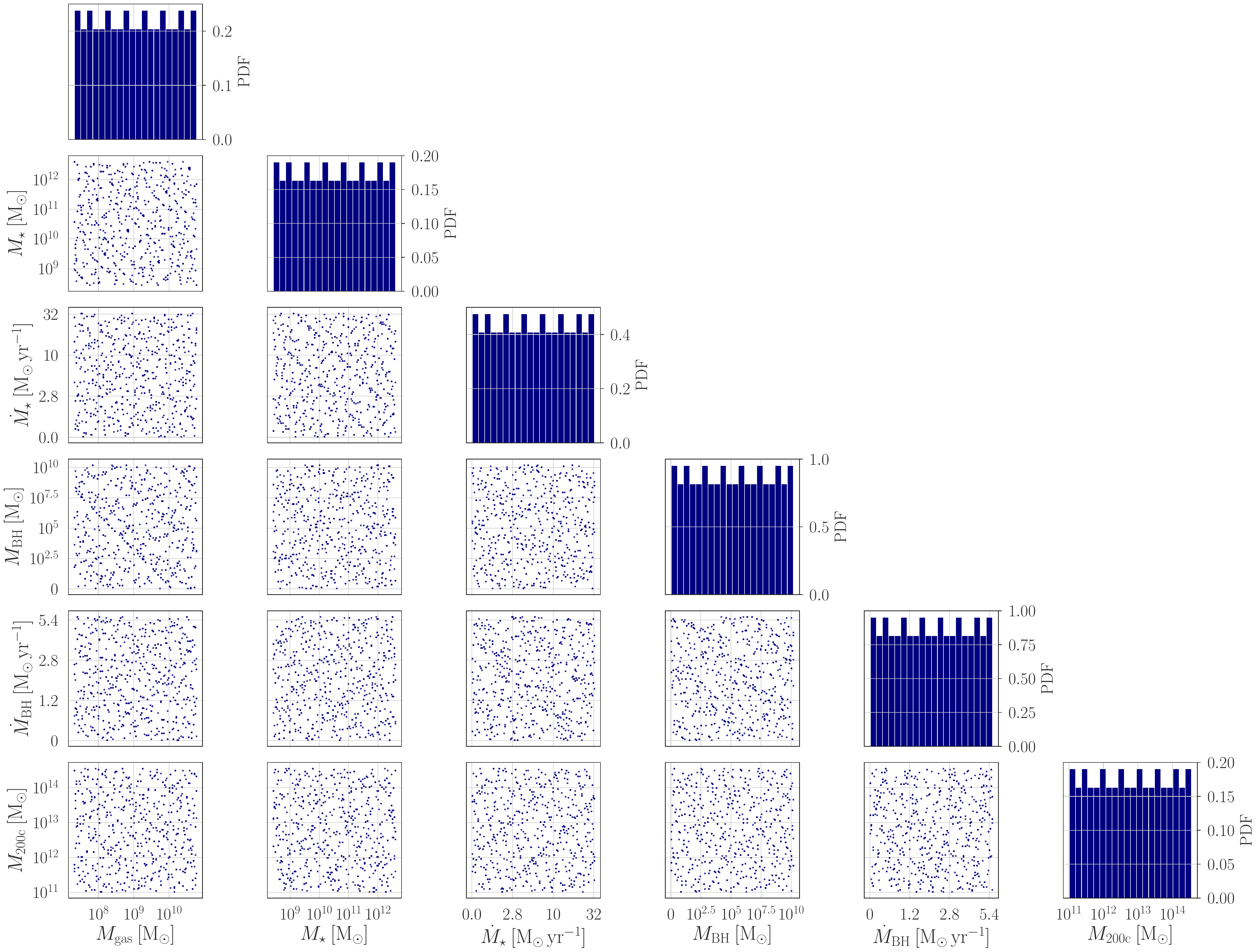}
    \caption{Sampling of the parameter space of the input features obtained from generating a Saltelli sequence \protect\citep{Saltelli_2010} This technique, combined with the modified inverse hyperbolic sine function applied to certain features ($\dot{M}_{\star}$, $\dot{M}_{\rm BH}$, $M_{\rm BH}$; see \S~\ref{sec:features} and Figure~\ref{fig:arcsinh}), ensures unbiased sampling over several orders of magnitudes, including null values. For illustrative purposes, this figure refers to a Saltelli sequence with index $N_S=2^7$, but we adopted $N_S=2^{14}$ in our analysis, to more finely sample the parameter space (see \S~\ref{sec:sampling} for details). In this figure, the boundaries each feature matches the ranges found for the Simba-100 simulation. We change them accordingly when we apply sensitivity analysis to other simulations. 
    }
    \label{fig:saltelli}
\end{figure*}

Sobol indices are useful to quantify the sensitivity of a model predicting a scalar quantity from a set of independent variables. In our case, the model is the RF algorithm described in \S~\ref{sec:algorithm}, and the independent variables the input features listed in \S~\ref{sec:features}. However, the predicted target variables are the gas densities at different radial distances from the halo centre. These constitute a data vector, therefore our model does not predict a scalar quantity. Nevertheless, for the sake of the sensitivity analysis, we can treat the predictions of the gas density in each radial bin as its own scalar model:
\begin{equation}
\label{eq:SA_model}
    \hat{\rho}(r_j) = \hat{\rho}_j (\, M_{\rm gas}, \, M_{\star}, \, \dot{M}_{\star}, \, M_{\rm BH}, \, \dot{M}_{\rm BH}, \, M_{\rm 200c}) \, ,
\end{equation}
where $\hat{\rho} (r_j)$ is the gas density at the $j$-th radial bin, as predicted by model $\hat{\rho}_j$. We will then derive the Sobol indices for each $\hat{\rho}_j$ model separately.

It is now convenient to define a data vector for our input features, $X=\{ X_k \} _{k=1,\dots,6}$, where each $X_k$ corresponds to one of the features in equation~\eqref{eq:SA_model}. If the input features are drawn from a given probability density function $p(X)$, we can straightforwardly calculate the expectation value, $\mathds{E}(\hat{\rho}_j)$, and variance, $\mathrm{Var}(\hat{\rho}_j)$, of the gas density profile in the $j$-th radial bin. By definition, these quantities refer to the response of the model to variations of all its input features over the entire parameter space. If, instead, we vary only one feature, say $X_{l}$, the corresponding variance on the gas density profile at each radial bin will be smaller. Precisely, the variance stemming from changing $X_{l}$ only is:
\begin{equation}
\label{eq:var_ord1}
    V_l^{(j)} = \int [E_l^{(j)} - \mathds{E}(\hat{\rho}_j)]^2 p(X) \, \mathrm{d}X_l\, ,
\end{equation}
where
\begin{equation}
\label{eq:exp_ord1}
    E_l^{(j)} = \mathds{E}(\hat{\rho}_j \vert X_k) = \int \hat{\rho}_j(X) p(X) \prod_{k \neq l} \mathrm{d}X_k\, 
\end{equation}
is the expectation value of the gas density in the $j$-th radial bin, conditional to feature $X_l$ being fixed, i.e., integrated over all features except for $X_l$.

The first-order Sobol index for the input feature $X_l$ is defined as
\begin{equation}
    \label{eq:S1_def}
    S_l^{(j)} = \frac{V_l^{(j)}}{\mathrm{Var}(\hat{\rho}_j)} \, .
\end{equation}
In other words, it is the contribution of changes in $X_l$ to the total variance on the predicted gas density at a given distance. If \smash{$S_l^{(j)}=1$}, then the total variance in $\mathrm{Var}(\hat{\rho}_j)$ is entirely determined by the input feature $X_l$. If \smash{$S_l^{(j)}=0$}, then variations of $X_l$, all other parameters being fixed, do not contribute to the total variance in $\mathrm{Var}(\hat{\rho}_j)$ at all. By construction, \smash{$S_l^{(j)}$} is confined within these limits, and in general it will assume some intermediate value. The value of \smash{$S_l^{(j)}$} can therefore be seen as quantifying the `importance' of feature $X_l$ in the model.

The first-order Sobol indices can then already provide great insight on the physical interpretation of the model. However, they do not bear information on the possible interactions amongst different input features. This is accounted for by higher-order Sobol indices. For example, one can consider the contribution to the total variance stemming from varying two features, say $X_l$ and $X_m$, at the same time:
\begin{equation}
\label{eq:var_ord2}
    V_{l m}^{(j)} = \int [E_{l m}^{(j)} - \mathds{E}(\hat{\rho}_j)]^2 p(X) \, \mathrm{d}X_l \, \mathrm{d}X_m - V_l^{(j)} - V_m^{(j)}  \, ,
\end{equation}
where
\begin{equation}
\label{eq:exp_ord2}
    E_{l m}^{(j)} = \mathds{E}(\hat{\rho}_j \vert X_l, \, X_m) = \int \hat{\rho}_j(X) p(X) \prod_{k \neq l,m} \mathrm{d}X_k \, . 
\end{equation}
Subtracting \smash{$V_l^{(j)}$} and \smash{$V_m^{(j)}$} from the integral in equation~\eqref{eq:var_ord2} ensures that the contributions to the total variance arising from changing features $X_l$ and $X_m$ individually is not double counted (since they already appear in the definition of the first-order Sobol indices). The corresponding second-order Sobol index is then simply \smash{$S_{l m}^{(j)} = V_{l m}/\mathrm{Var}(\hat{\rho}_j)$}. Following the same logic, it is possible to define any higher-order Sobol index, reflecting all possible combinations of three or more features. By construction, the sum of all possible Sobol indices must equate to unity. 

For each input feature $X_l$, it is convenient to introduce a total-order index that encapsulates all contributions of that feature to the total variance, at any order. Since in our case we have six input features, the total-order index is defined as
\begin{equation}
\label{eq:ST_def}
    S_{\mathrm{T}, \, l}^{(j)} = S_l^{(j)} + \sum_k S_{lk}^{(j)} + \sum_{k<n} S_{lkn}^{(j)} + \dots + S_{123456}^{(j)} \, .
\end{equation}
Thus, \smash{$S_{\mathrm{T}, \, l}^{(j)}$} tells us about the global response of the model to variations in feature $X_l$. It is also clear from equation~\eqref{eq:ST_def} that the difference between the total-order and first-order Sobol indices quantifies the interaction of the input feature in question with all other features.

To better understand how to interpret the first-order and total-order Sobol indices for a certain feature, it is instructive to consider three extreme cases. If \smash{$S_l^{(j)} = 0$} and \smash{$S_{\rm T, \, l}^{(j)} = 0$}, then the model does not depend on feature $l$ at all. Conversely, if \smash{$S_l^{(j)} =1$} and \smash{$S_{\rm T, \, l}^{(j)} = 1$}, then the mode depends on feature $l$ only. Finally, if \smash{$S_{\mathrm{T}, \, l}^{(j)} =1$} but \smash{$S_l^{(j)}=0$}, it means that the predictions of the model do not change when feature $l$ is varied \textit{individually}. However, varying \textit{other} features for some values of $X_l$ significantly affects the predictions of the model\footnote{An analytical example of this potentially counter-intuitive situation can be found \cite{Ishigami_1991}. For a summary of the corresponding Sobol sensitivity analysis, see \cite{Oleskiewicz_2020}.}.

Therefore, the first-order Sobol indices quantify the impact on the model predictions of each input feature when varied independently. If some input features are correlated, such as stellar mass and total halo mass in our case, that will be reflected in the higher-order indices. Thus, deriving the first-order and total-order Sobol indices enables us to obtain an overview of the importance and interaction of the input features in the model in question. We will focus on these indices in the remainder of the manuscript. We shall denote the list of first-order and total-order Sobol indices as $\mathbf{S}_1$ and $\mathbf{S}_{\rm T}$, respectively. In most practical applications, these indices need to be computed numerically. We will explain our procedure for this purpose in the next subsection.

\subsection{Numerical implementation}
\label{sec:sampling}

To compute Sobol indices in practice, we need to generate samples of input features from a certain probability density function $p(X)$ and then evaluate all integrals in equations~\eqref{eq:var_ord1}-\eqref{eq:exp_ord2}, as well as their higher-order analogues, numerically. It is important to choose $p(X)$ such that the parameter space of the input features is properly sampled and the resulting estimates of the Sobol indices are unbiased. 

A fast, well-converged, and fair sampling of the parameter space can be achieved by generating a so-called `Saltelli sequence' \citep{Saltelli_2010}, which constitutes a more efficient implementation of a previous sampling technique \citep{Sobol_1967}. To generate a Saltelli sequence, one needs to specify the boundaries of each input feature, and an index $N_S$, which we will call `Saltelli index'. For a parameter space with dimension $d$ (i.e., $d$ input features), the Saltelli sequence will contain $2 N_S(d+1)$ samples. We will not go into the details of the computational method to obtain a Saltelli sequence, referring the interested reader to the original publications \citep{Sobol_1967, Saltelli_2010}. Instead, we provide a visualisation of the resulting sample.

This is done in Figure~\ref{fig:saltelli}, where each off-diagonal panel in the corner plot shows the sample elements (plotted as dots) in the Saltelli sequence, projected onto a two-dimensional slice of the six-dimensional parameter space of the input features. The panels on the diagonal of the corner plot show the one-dimensional probability distribution function of each feature. Visually, the uniform pseudo-randomness of the sequence should be apparent. The boundaries of each feature in Figure~\ref{fig:saltelli} are defined by the full dataset extracted from the Simba-100 simulation. However, upon performing sensitivity analysis on other simulations, we change the boundaries accordingly. This ensures that we always explore all available parameter space. 

The importance of our rescaling in equation~\eqref{eq:archsinh} should now be evident. Indeed, it yields a fair sampling from zero up to arbitrarily large values of the input features. This is crucial for $\dot{M}_{\star}$, $\dot{M}_{\rm BH}$ and, most of all, $M_{\rm BH}$, which varies between zero and $10^{10} \Msun$. Had we generated a Saltelli sequence on a linear scale, the parameter space would have been disproportionally sampled at the high end of each input feature. Conversely, a logarithmic scale would have enhanced the sampling closer to the lower bound of each feature, while still excluding null values. Thus, in either case, the sampling would have been biased, effectively invalidating the subsequent sensitivity analysis.

For illustrative purposes, in Figure~\ref{fig:saltelli} we set $N_S = 2^7$, since for much higher indices the samples becomes so dense that individual dots are no longer distinguishable. But for our analysis we set $N_S=2^{14}$, which we verified to guarantee converged estimates for the Sobol indices. In practice, we generate the Saltelli sequence and perform our sensitivity analysis with the publicly available \texttt{SALib} Python package \citep{Salib}. This package returns the estimates of the first-order, second-order, and total-order Sobol indices, together with the $95\%$ confidence interval. 

The computation of the full set of Sobol indices can become prohibitively expensive as the number of input feature increases. For a parameter space with dimension $d$, there are in total $2^d-1$ Sobol indices. In our case, that would correspond to 63 indices. To reduce the computational cost, the \texttt{SALib} package neglects all Sobol indices beyond order two. This is well justified as long as the Sobol indices for all input features rapidly decrease as the order increases \citep{Salib}. We verified that this is indeed the case for our model, since the second-order indices are typically between one and two order of magnitudes smaller than their associated first-order ones. 

In our analysis, we will therefore focus on the $\mathbf{S}_1$ and $\mathbf{S}_{\rm T}$ indices only. \texttt{SALib} outputs their best estimates, together with their $95\%$ confidence intervals. In the next section, we will present the outcome of our sensitivity analysis for all radial bins, redshifts, and simulations. We will focus our discussion on the physical implications of the results.

\section{Results and discussion}
\label{sec:discussion}

\begin{figure*}
    \centering
    \includegraphics[width=\textwidth]{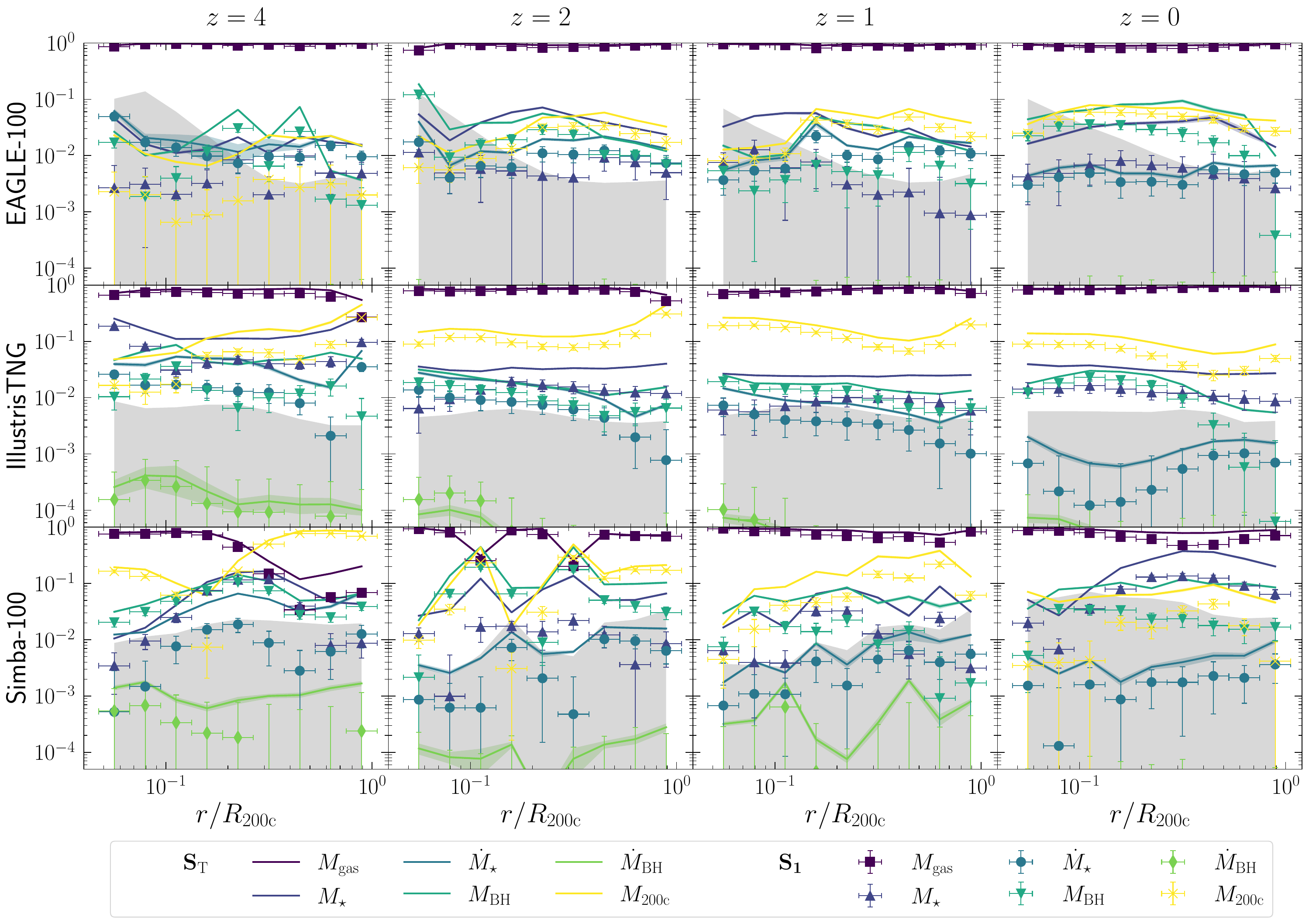}
    \caption{Results of the Sobol sensitivity analysis applied to our RF algorithm. Each row of panels refers to a different flagship simulation, as indicated in the left part of the figure. Every column corresponds to a different redshift, as reported in the top of the figure. Within each panel, the data points represent the $S_1$ Sobol coefficients representing the importance of a certain input feature in predicting the gas density profile at different radial bins. Each feature is associated with a different colour and marker style, as indicated in the legend. The error bars report the 95\% confidence level associated with the $S_1$ coefficients at each radial bin. In every panel, the solid lines represent the $S_{\rm T}$ Sobol coefficients, and the shaded areas the corresponding 95\% confidence level. The colour coding for the $S_{\rm T}$ coefficients is the same as for the $S_1$ coefficients. The grey shaded area in every panels represents the region within which the variance on the predicted density profile introduced by any input feature would be smaller than the intrinsic accuracy of the gas density profile predicted through our RF algorithm. Thus, input features with Sobol indices falling below the threshold between the grey and white regions do not have a physically meaningful effect on the gas density profile at the radial bin considered. In all simulations and at all redshifts, the total gas mass of the central galaxy is the most impactful feature on the prediction of the as density profile. The SFR and the accretion rate of the BH in the central galaxy are the least consequential properties.
    }
    \label{fig:sobol_z}
\end{figure*}

In this section, we present the results of the sensitivity analysis of our model. We will begin with comparing the relative importance of different input features in the flagship simulations EAGLE-100, IllustrisTNG, and Simba-100 (\S~\ref{sec:importance}). Then, we will focus on the other Simba runs, studying how feature importance changes under the different feedback variants (\S~\ref{sec:feedback}). Finally, in \S~\ref{sec:limitations}, we will discuss the limitations of our model and prospects for improvement.

It is important to note that the input features of our model under a given feedback scenario are not independent input parameters of the corresponding simulation. In fact, the halo mass and the properties of the central galaxy that we consider are themselves predictions of the simulation, in the same way as the gas density profiles are. Thus, whenever we say that a certain feature has `high importance' for, or is a `strong predictor' of the gas density profile, we do not mean that this is necessarily a cause-effect relationship. Rather, the implication is that the galaxy formation model embedded in the specific simulation considered generates strong correlations between the feature in question and the corresponding gas density profile. 

We could speak about causation if we ran our sensitivity analysis on the parameters of the feedback prescription within each simulation. We chose not to do that because they are usually code-specific numerical parameters, which are not always physically well defined. Instead, the main motivation of our work is to understand the physics that shapes the gas density distribution in haloes with different properties. That is the reason why we built our model on physically grounded input features. As such, the physical implications of our model are the main focus of this section.

\subsection{Flagship simulations}
\label{sec:importance}

Figure~\ref{fig:sobol_z} shows the results of the sensitivity analysis that we performed on our model for the flagship simulations EAGLE-100, IllustrisTNG, and Simba-100, each occupying a distinct row of panels. The vertical panels refer to the same redshift, as indicated in the upper part of the figure, so that we can also appreciate the redshift evolution of the Sobol indices. The data points in each panel refer to the best estimates of the first-order indices, plotted as a function of the radial distance from the centre of the halo. The position of the data point lies in the middle of the logarithmic radial bin, and the horizontal error bars represent the bin width. The vertical error bars span the 95\% confidence interval on the estimate of the Sobol index. The colour and marker style of the data points reveal the input features that the indices in question are referring to (see legend). The solid lines represent the best estimate of the total-order indices as a function of the radial distance, following the same colour coding. The shaded area around each solid line spans the 95\% confidence interval on the estimate. 

From the logarithmic scale of the $y$-axis in every panel, we immediately notice that impact of different features on our model can vary greatly, from being consistent with zero to coming close to unity. It is tempting to immediately discuss the physical implications of the rank ordering of the Sobol indices for each input feature in the various simulations and redshifts considered. However, we should be wary of over-interpreting the relative impact of features associated with low Sobol indices. The key question is then what the minimum threshold for deeming a Sobol index as `physically significant' is. To answer this question, we should go back to the definition of Sobol indices, starting from order one.

\subsubsection{Significance threshold}
\label{sec:threshold}

Equation~\eqref{eq:S1_def} tells us that the first-order Sobol index of an input feature is the fractional contribution to the total variance of the model that stems from varying that feature individually. If this is smaller than the intrinsic accuracy of the model, then the importance of the associated input feature is statistically insignificant. Precisely, we require that
\begin{equation}
    \sqrt{V_l^{(j)}} \gtrsim \frac{1}{N_{\rm h}} \sum_{i=1}^{N_{\rm h}} \vert \log \hat{\rho}_{ij} - \log \rho_{ij} \vert \, := 1 - \bar{\alpha}(r_j) \;,
\end{equation}
where the quantity on the left hand side is the root mean square associated with the variance \smash{$V_l^{(j)}$}, as defined in equation~\eqref{eq:var_ord1}. The equality at the right hand side descends by definition from equation~\eqref{eq:accuracy}, with the bar highlighting that we are averaging the accuracy $\alpha(r_j)$ over all haloes. The accuracy is evaluated with respect to the training set now: if the response of our algorithm to the variation of a given input feature falls below even the best theoretically possible accuracy, then surely such response should not be deemed significant.

Recalling the definition given in equation~\eqref{eq:S1_def}, it follows that the first-order Sobol indices of all features relative to halo-centric distance $r_j$ must exceed the threshold
\begin{equation}
\label{eq:threshold}
    S_{\rm th}^{(j)} = \frac{[1- \bar{\alpha}(r_j) ]^2}{ \mathrm{Var}({\hat{\rho}_j}) } \, ,
\end{equation}
in order to be considered physically significant in the context of our model. The variance in the denominator of the equation above refers to the total variance of our model as introduced in \S~\ref{sec:sensitivity}. While for consistency of notation we wrote $\hat{\rho}_j$ instead of $\log \hat{\rho}_j$, we remind the reader that our target variables are always (within good approximation -- see \S~\ref{sec:training}) the logarithm of the gas density profile, and not the gas density profile itself.

From equation~\eqref{eq:threshold}, we can see that a higher accuracy therefore diminishes \smash{\smash{$S_{\rm th}^{(j)}$}}, since smaller variations of the input feature examined will become statistically significant. Similarly, a higher intrinsic variance decreases the threshold, because the variance introduced by changing the input feature in question will become smaller in relative terms. Thus, a smaller Sobol index can become physically significant.

Since the total-order Sobol index of a feature cannot be lower than the associated first-order index, the threshold defined in equation~\eqref{eq:threshold} applies to $\mathbf{S}_1$ and $\mathbf{S}_{\rm T}$ alike. The value of \smash{$S_{\rm th}^{(j)}$} for each snapshot of every flagship simulation corresponds to the line partitioning the corresponding panel in Figure~\ref{fig:sobol_z} between the white and grey shaded areas. Thus, Sobol indices within the grey area are not considered to be physically significant for our purposes. Those consistent with \smash{$S_{\rm th}^{(j)}$} within the $95\%$ confidence level are only marginally significant.

IllustrisTNG and Simba-100 exhibits thresholds between $0.002$ and $0.2$ over all radial bins. At $z=0$, Simba-100 is characterised by a higher threshold ($0.03-0.04$) within $0.1 \, R_{\rm 200c}$. This reflects the lower accuracy achieved closer to the centre for some haloes, as seen in Figure~\ref{fig:2D_dens_prof}. Similarly, the model generally performs less well at lower radial distances in EAGLE-100. This translates into an even more evident rise in the Sobol threshold within $0.2 \, R_{\rm 200c}$ at all redshifts.

\subsubsection{Feature importance}

We are now ready to discuss the results of the sensitivity analysis shown in Figure~\ref{fig:sobol_z}. A striking outcome is that the gas mass of the central galaxy is the most important feature to predict the gas density profile at all redshifts, and in all simulations. In most cases, the first-order and total-order Sobol indices for $M_{\rm gas}$ are roughly equal, implying that this property does not interact appreciably with the other features. Thus, $M_{\rm gas}$ single-handedly imprints the largest variations on the gas distribution. This is not surprising: because of the steep radial decrease of the gas density profile, a sizeable fraction of the gas mass within the halo is confined within $0.05 - 0.1 \, R_{\rm 200c}$. The central galaxy typically resides within this region, and contains a significant share of the enclosed gas mass. Thus, $M_{\rm gas}$ is expected to correlate with the gas mass within the halo. The latter has clear impact on the gas density profile, as it sets its normalisation. Thus, because of the correlation between $M_{\rm gas}$ and the halo gas mass, it is physically reasonable to expect that this feature has an important impact on the predicted gas density profile.

The only exception is represented by Simba-100 at $z=4$. In this case, $M_{\rm gas}$ is the most important feature only within $\sim 0.2 \, R_{\rm 200c}$. There are two reasons why this happens. First, haloes are denser at higher redshift. For a given gas mass, the gas is therefore more concentrated towards the halo centre \citep[see e.g.][]{Sorini_2025}. Secondly, the cutoff of the halo mass function is shifted towards lower masses at higher redshift. Therefore, the results at $z=4$ are biased towards lower halo masses. This means that haloes contain, on average, fewer gas elements, which are more concentrated towards the centre. This is especially relevant for Simba-100, which has the lowest mass resolution among the flagship simulations considered -- meaning even fewer gas elements in a halo of a given mass. The end result is that the gas mass matters only at smaller halo-centric distances; beyond $\sim 0.2 \, R_{\rm 200c}$ there are simply not enough gas resolution elements to make a statistically meaningful impact on the gas density profile. There are plenty of DM particles, though. Understandably, beyond the aforementioned radius, the halo mass $M_{\rm 200c}$ takes over as the most impactful feature. 

The halo mass is the second most important feature in all simulations. This should not be too surprising either, as $M_{\rm 200c}$ indicates the type of object that is being considered: dwarf, Milky-Way-like galaxy, group, cluster. Each class is known to exhibit different relative strengths of astrophysical processes such as star formation, stellar and AGN feedback \citep[see e.g.][]{Crain_2023}. Accordingly, $M_{\rm 200c}$ acts as a global predictor of the gas density profile: its Sobol indices are generally only weakly dependent on the halo-centric distance. Unlike $M_{\rm gas}$, though, the difference between the total-order and first-order indices is never negligible. This reflects the strong correlations between the halo mass with other properties of the central galaxy. 

The importance of $M_{\rm 200c}$ is particularly evident in the case of IllustrisTNG at $z\leq 2$, where the Sobol indices of all input features other than $M_{\rm gas}$ are at least $0.8 \, \rm dex$ smaller than those of $M_{\rm 200c}$. Thus, properties other than gas and halo mass have a sub-dominant impact on the predictions of our model. However, at $z=4$, the first-order Sobol indices for $M_{\star}$ are above \smash{$S_{\rm th}^{(j)}$}, contributing to the total variance of the gas density within $\sim 0.1 \, R_{\rm 200c}$ to a greater extent than $M_{\rm 200c}$. At such high redshift, star formation and stellar feedback are the dominant sub-grid processes \citep{Weinberger_2017}, thus $M_{\star}$ is a strong predictor of the gas distribution close to the central galaxy. The total-order index for this property is well above the significance threshold, underscoring the importance of the interaction of the stellar mass with the other input features.

In the EAGLE-100 simulation, the situation at $z=4$ is somewhat atypical. The first-order Sobol indices of the halo mass are below the significance threshold. Again, this could be a reflection of the bias towards lower-mass haloes at higher redshift. With no galaxy groups or clusters in the simulation, $M_{\rm 200c}$ loses the discriminatory power among different types of objects that it enjoys at lower redshift, where the dynamic range is wider. The SFR, BH and (to a lesser extent) stellar mass become more important in defining the gas distribution, as these are indicators of the strength of star formation and feedback within central galaxies residing in haloes of similar mass. Although less impactful than $M_{\rm 200c}$, the SFR and BH mass imprint a statistically significant impact at $z \leq 2$ too. In particular, at $z=0$ the relative importance of the BH mass dramatically increases with respect to that of the SFR and stellar mass. This is a consequence of the higher share of haloes undergoing AGN feedback, the energy output of which is set by the BH mass (see \S~\ref{sec:EAGLE}). 

We observe something similar in Simba-100 at $z=0$, where halo mass, BH and stellar mass have comparable importance. The large differences between the corresponding total-order and first-order Sobol indices highlight the strong interaction of these properties with all other input features. The Sobol indices of the stellar mass are above the significance threshold for bins beyond $\sim 0.1 R_{\rm 200c}$, with the total Sobol indices being particularly high. Thus, the model responds significantly to the stellar mass both directly and through its correlations with the other galaxy and halo properties. Correlations between the star formation history in the Simba suite of simulations and BH properties have been extensively studied and connected to the underlying feedback prescriptions (see \citealt{Thomas_2019}, \citealt{Scharre_2024}, and \S~\ref{sec:feedback}). At $z=1$, the model still responds significantly to $M_{\rm BH}$ and $M_{\star}$, mainly through the interactions with other properties. At $z=2$, the first-order Sobol indices for the BH mass occasionally increase to the same level of those for the gas mass and halo mass. Interestingly, $z=2$ is the redshift at which AGN feedback in Simba-100 begins to impact significantly both the gas distribution of gas at all scales, and the star formation history \citep{Sorini_2022, Scharre_2024}. That explains the higher Sobol index of $M_{\rm BH}$ at this redshift.

The BH accretion rate has no significant impact on the predictions of our model at any redshift nor in any simulation. Recalling that our input feature is the \textit{instantaneous} accretion rate, this could in part reflect a timescale issue. Whereas the BH accretion rate varies on galactic timescales, the CGM responds on halo timescales. Thus, when viewed at a given instant, the BH accretion rate tends to exhibit a weaker coupling with the gas density profile than its time-integrated counterpart (i.e., the BH mass).

Similar considerations would  in principle apply to the SFR and stellar mass, although we do not observe the same relative difference in the importance of these quantities as in the case of the BH accretion rate and mass. As a further test, we tried replacing the instantaneous BH accretion rate with its average over five snapshots. However, upon re-training our models, the outcome of the sensitivity analysis did not change. This suggests that the BH accretion rate has a genuinely low impact on the estimation of the gas density profile, regardless of the feedback model, for reasons that might go beyond the timescales considerations illustrated earlier. At first sight, this might seem puzzling, since the BH accretion rate does play a crucial role in the BH growth and AGN feedback prescriptions of all simulations. Since simulations indicate that higher BH accretion rates are generally correlated with higher baryon mass fraction within haloes (Denison et al., in prep.), one might expect a stronger connection with the gas density profile as well, at least in the inner regions of the halo.

We have two possible explanations for our somewhat counter-intuitive result, not necessarily mutually exclusive. First, the gas density profile is a volume averaged, spherically symmetric, quantity. If a certain AGN feedback prescription, such as the AGN-wind and AGN-jet modes in Simba, is highly anisotropic, it is likely to affect the gas distribution mostly along the direction of the angular momentum of the accretion disk \citep{Yang_2024}. But the affected region occupies a small volume fraction within the halo boundary, hence the signature of AGN feedback on the gas density profile will be diluted. Secondly, the impact of AGN feedback on the gas distribution can extend out to several virial radii, whether the prescription is isotropic or not (\citealt{Sorini_2022, Ayromlou_2023, Ondaro-Mallea_2025}). Since we restrict ourselves to the region within the virial radius, that can explain why we do not observe any significant signature of the BH accretion rate.

We verified that the rank ordering of the Sobol indices associated with the different input features remains consistent under variations of the box size and mass resolution for all simulations (see Appendix~\ref{app:robustness} for details). This validates the robustness of our physical interpretation of the sensitivity analysis results presented in this section.

\subsection{Feedback variants}
\label{sec:feedback}

\begin{figure*}
    \centering
    \includegraphics[width=\textwidth]{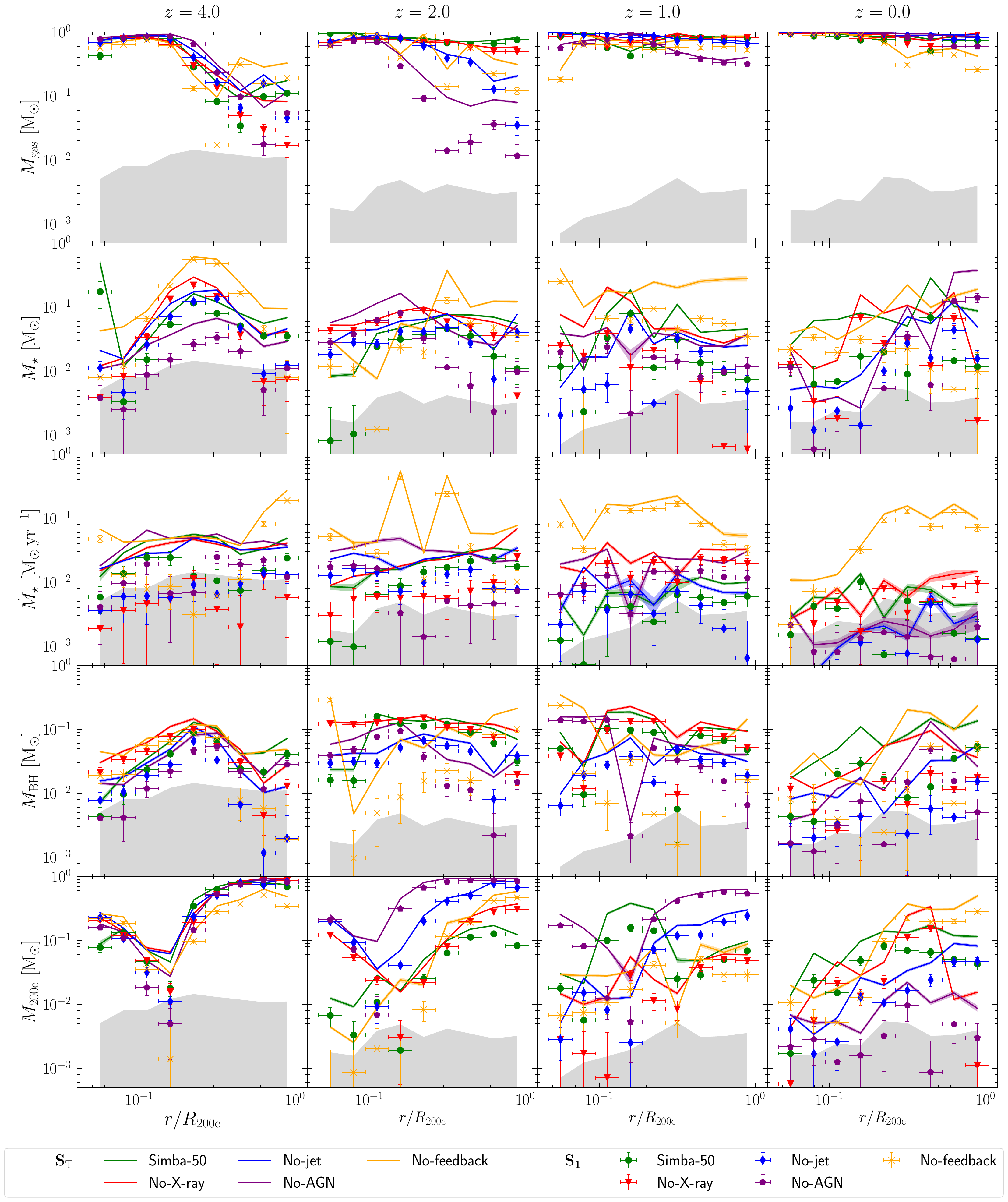}
    \caption{Results of the Sobol sensitivity analysis for the variants of the Simba simulation. Every row of panels corresponds to a different input feature, as indicated in the left part of the figure. Each column of panels is associated with a different redshift, as reported in the top part of the figure. The data points and solid lines represent the $S_1$ and $S_{\rm T}$ Sobol coefficients, as a function of the radial distance, for different runs (see legend). The meaning of the error bars and shaded areas are the same as in Figure~\ref{fig:sobol_z}. In all runs, the gas mass and $M_{\rm 200c}$ are the strongest predictors of the gas density profile. Stellar feedback reduces the overall impact of the SFR of the central galaxy, and the activation of AGN feedback mechanisms increases the importance of the BH and stellar mass in shaping the gas density distribution.
    }
    \label{fig:sobol_simba}
\end{figure*}

From the previous section, it is clear that sensitivity analysis can provide insightful information on the physics connecting galaxy and halo properties with the corresponding gas density profile. Certain outcomes, such as the global importance of $M_{\rm gas}$ and $M_{\rm 200c}$ are nearly universal, regardless of the galaxy formation model embedded in each simulation. On the other hand, properties such as the SFR, stellar or BH mass impact the gas density profile to different extents depending on the underlying feedback prescriptions. While in \S~\ref{sec:importance} we compared simulations with qualitatively different feedback models, we will now consider variations of the same feedback scheme. Taking advantage of the Simba $50 h^{-1} \cMpc$ runs listed in Table~\ref{tab:simulations}, we can turn off individual feedback modules and meticulously study the consequences on the connections between galaxy and halo properties and the surrounding gas density profile.

We repeat the Sobol sensitivity analysis explained in \S~\ref{sec:sensitivity} on all Simba feedback variants and present the results in Figure~\ref{fig:sobol_simba}. Panels belonging to the same column refer to the same redshift. Unlike Figure~\ref{fig:sobol_z}, each row of panels now corresponds to a specific input feature. We did not include the BH accretion rate, because its Sobol indices are always consistent with zero, as discussed in \S~\ref{sec:importance}. In each panel, the first-order and total-order Sobol indices are represented with data points and solid lines, respectively, with colours (and, if applicable, marker styles) identifying different Simba variants. The meaning of the error bars and shaded areas are the same as in Figure~\ref{fig:sobol_z}. We calculate the significance threshold \smash{$S_{\rm th}^{(j)}$} as in equation~\eqref{eq:threshold} for each simulation. To not overcrowd the plot, we chose to draw the shaded grey area below the maximum \smash{$S_{\rm th}^{(j)}$} obtained across all runs for any given redshift.

As already found in \S~\ref{sec:importance}, at $z=4$ the gas mass of the central galaxy impacts the gas density profile most strongly within $0.1 \, R_{\rm 200c}$. This is to be expected, since at high redshift galaxies have had less time to eject their gas far into their CGM, thus $M_{\rm gas}$ is a stronger predictor of the gas density profile near the central galaxy only. Beyond this distance, the gas distribution is primarily shaped by the overall distribution of DM and baryons within the halo instead, hence the higher importance of $M_{\rm 200c}$. This is true for all feedback variants, suggesting that the importance of $M_{\rm gas}$ and $M_{\rm 200c}$ is not tied to the specifics of subgrid physics. 
 
At $z=2$, the importance of $M_{\rm gas}$ extends all the way to the virial radius in the runs that include at least the AGN-jet mode. This happens because a higher gas mass can feed the central BH more effectively, pushing it beyond the mass and accretion rate thresholds that activate the high-speed, far-reaching AGN-jet mode in the Simba simulation. On the contrary, the other feedback variants still exhibit a lower impact of $M_{\rm gas}$ farther from the central galaxy. Conversely, the $M_{\rm 200c}$ impact beyond $\sim 0.2 \, R_{\rm 200c}$ at $z=2$ is stronger for the runs without AGN-jets. 

At lower redshifts, as all feedback modes have had more time to influence the CGM of the central galaxy, $M_{\rm gas}$ progressively becomes the most important predictor of the gas density profile at all radial bins considered, whereas the relevance of $M_{\rm 200c}$ is somewhat diminished, especially within $\sim 0.2 \, R_{\rm 200c}$. As expected, the No-feedback run constitutes an exception, in that it shows declining importance of $M_{\rm gas}$, and rising weight of $M_{\rm 200c}$, farther from the central galaxy. Indeed, in the absence of any feedback mode, it becomes harder for the central galaxy to eject its original gas content towards the outskirts of the halo. This explains the results of the No-feedback run compared to all other variants, but it might still look puzzling that even without any feedback process, the impact of $M_{\rm gas}$ reaches farther distances by $z=0$. The reason is that gas can still be transmuted into stars, and this process is particularly efficient in the absence of any self-regulating mechanisms. Star formation happens not only in the central galaxy, but also in the satellites, which can be spread all over the host halo. Thus, we can still detect a correlation between $M_{\rm gas}$ and the gas density distribution at the outskirts of the halo even in the No-feedback run.
    
The BH mass of the central galaxy has roughly the same impact in all runs at $z=4$. At later redshifts, the effect of $M_{\rm BH}$ on the gas density profile is more important for the runs that contain at least the AGN-jet mode. This reflects the fact that the BH mass is one of the key parameters that activate the AGN-wind and AGN-jet modes. Since AGN-jets have higher speed, they can reach out to the outer regions of the halo more easily than AGN-winds. That is the reason why the No-X-ray and Simba-50 runs exhibit higher $\mathbf{S}_1$ coefficients than the No-jet run, especially at larger distances from the halo centre. In the simulations with at least AGN-winds, the importance of $M_{\rm BH}$ is slightly decaying with the radial distance at $z=2$, while at later redshifts the trend is progressively inverted. This follows from the fact that at later times AGN-winds and AGN-jets have had more time to pervade the entirety of the halo (and beyond; see e.g. \citealt{Sorini_2022}). The importance of $M_{\rm BH}$ in the No-AGN and No-feedback runs is mostly significant in the region near the central galaxy ($\lesssim 0.2 \, R_{\rm 200c}$). This does not mean that the BH mass impacts the gas density distribution by \textit{directly} influencing the strength of stellar feedback (in the case of the No-AGN run) or the gas properties. Rather, the BH mass is correlated with the stellar mass of the central galaxy, which in turn more closely affects the amount of gas in its environment through star formation and stellar feedback \citep{Thomas_2019, Scharre_2024}.

This is the reason why, in general, the trends of the Sobol coefficients for the stellar mass of the central galaxy is qualitatively similar to those already discussed for the BH mass. The importance of $M_{\star}$ at $z=4$ is higher in the No-feedback variant, which produces overall more stars as a consequence of the lack of feedback mechanisms. In the outermost radial bins, this is also the case for the SFR, which in the No-feedback run is almost as important as the gas mass of the central galaxy in predicting the gas density profile. The large difference with the corresponding total-order Sobol index underscores the strong interaction between SFR and the other input features in shaping the gas density profile. On the other hand, most first-order Sobol indices lie below the significance threshold, hence their impact is mostly negligible at $z=4$. Thereafter, the impact of the SFR acquires significance in shaping the gas density profile at all distances, especially for the No-feedback run, albeit with less importance than other input features such as $M_{\rm gas}$ or $M_{\rm 200c}$. This descends from the fact that since cosmic noon, newly formed stars have had more time to undergo stellar feedback processes, hence directly impacting the surrounding gas.

In conclusion, the gas mass and $M_{\rm 200c}$ remain principal predictors of the gas density profile within haloes for all feedback variants. However, the introduction of stellar feedback reduces the overall importance of the SFR of the central galaxy. Activating further AGN feedback mechanisms progressively increases the weight of the BH and stellar mass in shaping the gas density distribution.

\subsection{Limitations of this work}
\label{sec:limitations}

We have shown that a data-driven algorithm as simple as a RF can be trained on cosmological hydrodynamical simulations to accurately predict the halo gas density profile from the knowledge of a limited number of input features. Applying sensitivity analysis then allows us to physically interpret the predictions of our model, providing us with further insight on how different feedback scenarios influence the connection between galaxy and halo properties, and the surrounding gas density profile. It is remarkable that such a computationally cheap and physically intuitive approach can achieve a mean accuracy in the range $83-90\%$ in the redshift interval $0 < z < 4$ (see Figure~\ref{fig:accuracy_z}). However, this should be seen as a first milestone in a longer-term effort to understand the interplay between feedback, galaxy properties, and gas distribution. As such, our work can be ameliorated along different lines, which we shall pursue in future work.

There is certainly room for improving the accuracy of the model at $z<1$, which currently decreases for all simulations considered, and especially for the Simba suite. One way to improve performance may be to boost the statistics of rarer, higher-mass haloes, by expanding the training set with data from larger boxes (e.g. MillenniumTNG, \citealt{Pakmor_2023}, and FLAMINGO, \citealt{Flamingo}, for IllustrisTNG and EAGLE-like models, respectively) or from dedicated zoom-in campaigns targeting galaxy galaxy groups (e.g., HYENAS, \citealt{Hyenas}, for the Simba model). Even then, it might be that our algorithm as it stands may be too simplistic at $z=0$ and we may need to include additional input features. For example, the average gas mass ejected within feedback-driven outflows from a halo in a certain time scale may carry more direct information on both feedback activity and the gas distribution. Furthermore, extending the set of input features with analogous quantities from the satellite galaxies may improve the accuracy in the radial bins where such objects are located. More broadly, one may account for environmental variables as well, such as the cosmic web region (i.e., filament, knot, etc.) where the halo is located.

Obviously, there is a balance to be struck. Adding further input features may well improve the accuracy of the model, but would also increase complexity, hence inflating computational costs and risking overfitting. If, additionally, one plans to eventually adapt our model to constrain feedback prescriptions from observations of the gas distribution around galaxies, then the input features should preferably match well-defined observables. However, applications on observations are still premature. We would first need to extend the model such that the target variables include the density profiles of several gaseous phases, and not merely the total gas density profile. We should then forward model the profiles of a certain phase to obtain mock absorption or emission maps of the relevant tracers, to be compared with specific observational data. Again, we regard this manuscript as the first step towards this ambitious goal.

In this work, we considered simulations with very similar cosmological parameters. But in principle, it would be interesting to quantify any possible degeneracy between cosmology and galactic astrophysics in shaping the gas density profile. In practice, one could include the cosmological parameters in the set of input features and then perform the sensitivity analysis presented here. This would require training the algorithm on a multitude of feedback models and cosmologies. We plan to do this in the near future using the CAMELS suite of cosmological hydrodynamical simulations, which would provide an ideal test bed for this purpose.

\section{Conclusions and perspectives}
\label{sec:conclusions}

We studied the connection between galaxy properties and the radial gas density profile in their host haloes in the redshift range $0< z < 4$, in cosmological hydrodynamical simulations with different stellar and AGN feedback scenarios. We considered the flagship runs of the EAGLE, IllustrisTNG, and Simba simulations, as well as five variants of Simba where different subsets of the feedback modes had been deactivated. In all simulations, we extracted all well resolved haloes hosting at least one galaxy, and computed the spherically averaged total gas density profile as a function of the radial distance from the halo centre. For each simulation, we then trained a random forest algorithm predicting the halo gas density profiles from six input features: the total halo mass $M_{\rm 200c}$; the SFR, gas and stellar mass of the central galaxy; the mass and accretion rate of its central BH. To physically interpret our data-driven model, we performed a Sobol sensitivity analysis to quantify the impact of each input feature on the predicted gas density at different halo-centric distances, redshifts, and under different feedback scenarios. To the best of our knowledge, this is the first time that such statistical technique is applied on full cosmological hydrodynamical simulations. Our main conclusions are as follows: 

\begin{enumerate}
    \item A simple and efficient random forest algorithm can successfully predict the radial gas density profile of haloes, from the knowledge of just five global properties of the central galaxy, and the halo mass (Figures~\ref{fig:dens_prof}\&\ref{fig:2D_dens_prof}). This is true regardless of the specific feedback models. Our RF algorithm achieves an average accuracy of nearly 90\% for EAGLE and IllustrisTNG in the redshift interval $1<z<4$, and around 86\% at $z=0$. The Simba simulations exhibit slightly lower accuracy in the same mass and redshift ranges -- 83-84\% at $z=0$, and up to 87\% at higher redshift (Figure~\ref{fig:accuracy_z}).

    \item The results of our sensitivity analysis shows that for all simulations considered, the halo mass and the gas mass of the central galaxy are the most impactful properties that shape the gas density profile, at all redshifts considered, and at all distances from the centre of the halo (Figure~\ref{fig:sobol_z}). 

    \item The stellar mass of the central galaxy appears to be more important at higher redshifts, whereas the impact of the mass of the central BH increases at lower redshift (Figure~\ref{fig:sobol_z}). This is consistent with previous studies \citep[e.g.][]{Weinberger_2017, Scharre_2024} showing that stellar feedback is dominant at earlier times, whereas AGN feedback sets on at later times. In the case of Simba, the BH mass tends to be more important from $z=2$ onwards, and the stellar mass does not significantly impact the accuracy of the algorithm.

    \item The impact of the SFR of the central galaxy is sub-dominant with respect to the other properties considered. Also, in all simulations and redshifts analysed, the BH accretion rate does not appreciably impact the predictions of our model. Thus, its effect is either limited within the central galaxy, or it impacts a small volume fraction of the gas enclosed within the halo, hence not contributing appreciably to the overall gas density profile (Figure~\ref{fig:sobol_z}). 
    
    \item Comparing the results of Sobol sensitivity analysis across the different feedback variants of the Simba simulation, we found that the gas mass and $M_{\rm 200c}$ consistently rank as the strongest predictors of the gas density profile within haloes. However, the introduction of stellar feedback reduces the overall importance of the SFR of the central galaxy. Activating further AGN feedback mechanisms progressively increases the weight of the BH and stellar mass in shaping the gas density distribution (Figure~\ref{fig:sobol_simba}). 
\end{enumerate}

We stress that the input features are not parameters of the feedback model underlying each simulation, but predictions of the simulations themselves. Therefore, our sensitivity analysis is focused on exploring the connections between physically well defined galaxy or halo properties and the surrounding gas distribution, rather than on determining the importance of numerical feedback parameters in shaping the gaseous environment of galaxies. Each feedback model impacts both the input features and the target variables (i.e., the gas density profile), and therefore the detailed conclusions of the Sobol analysis. Nevertheless, some of our conclusions on the importance of different features apply across all simulations.

As it stands, our model can be readily included in analytical or semi-analytical models  \citep[e.g.][]{Lacey_2016, SP21, Pandya_2023} of galaxy formation to connect galaxy and halo properties with the surrounding gas density profile. In forthcoming work, we plan to develop analogous data-driven algorithms, and combine them with sensitivity analysis, to understand the impact of feedback on the radial profiles of different gaseous \textit{phases} (e.g., cold ISM, cool/warm/hot CGM). This would serve as a starting point for connecting feedback parameters in simulations with observational data tracking several gaseous phases around haloes of different masses and at different redshifts. Indeed, existing absorption \citep[e.g.][]{Prochaska_2013, Galbiati_2024} and emission-line observations \citep[e.g.][]{Zhang_2024}, together with ongoing and future large-scale surveys \citep[e.g.][]{WEAVE}, are mapping different gaseous environments, offering opportunities to jointly constrain the relevant feedback physics. With this longer-term aim in mind, we note that the accuracy of our algorithm is promising for this purpose (see \S~\ref{sec:validation}). Therefore, our work represents a successful proof of concept over which to construct a more refined physically interpretable ML model for understanding and constraining feedback models from the observed distribution of gas around haloes over cosmic time.



\section*{Acknowledgements}

DS and SB acknowledge funding from a UK Research \& Innovation (UKRI) Future Leaders Fellowship [grant numbers MR/V023381/1 and UKRI2044]. We acknowledge the \texttt{yt} team for development and support of \texttt{yt}. This work used the DiRAC@Durham facility managed by the Institute for Computational Cosmology on behalf of the STFC DiRAC HPC Facility, with equipment funded by BEIS capital funding via STFC capital grants ST/K00042X/1, ST/P002293/1, ST/R002371/1 and ST/S002502/1, Durham University and STFC operations grant ST/R000832/1. DiRAC is part of the National e-Infrastructure. This work made extensive use of the NASA Astrophysics Data System and of the astro-ph preprint archive at arXiv.org. For the purpose of open access, the author has applied a Creative Commons Attribution (CC BY) licence to any Author Accepted Manuscript version arising from this submission.

\section*{Data Availability}

The simulation data of all runs are publicly available\footnote{\url{http://simba.roe.ac.uk}}$^{,}$\footnote{\url{https://www.tng-project.org}}$^{,}$\footnote{\url{https://icc.dur.ac.uk/EAGLE/database.php}}. The derived data underlying this article will be shared upon reasonable request to the corresponding author.




\bibliographystyle{mnras}
\bibliography{gas_dens_prof_ml}



\appendix

\section{Robustness of sensitivity analysis to simulation volume and mass resolution}
\label{app:robustness}

\begin{figure}
    \centering
    \includegraphics[width=\columnwidth]{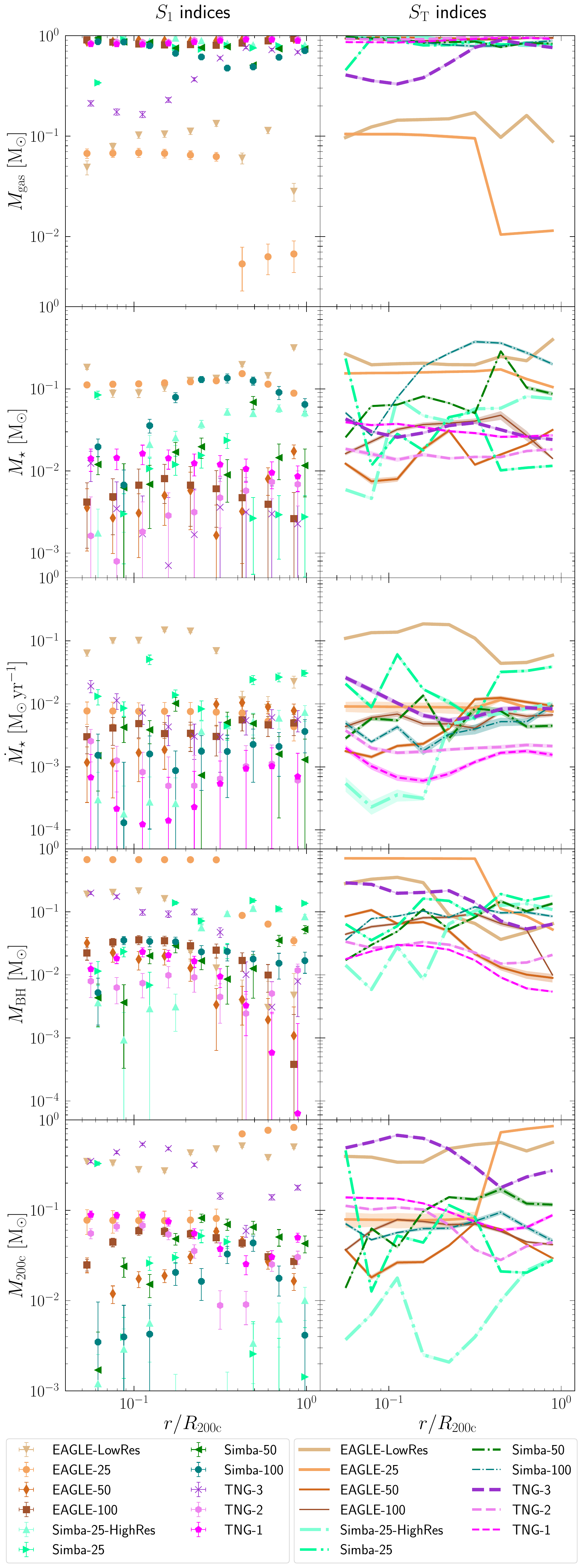}
    \caption{Convergence test with respect to mass resolution and box size at $z=0$ for all simulations considered. Each row refers to a different input feature and left and right panels report the first-order and total-order Sobol indices, respectively. There is good convergence volume wise but not with respect to mass resolution for EAGLE and Simba. Nevertheless, the relative importance of different input features in shaping the gas density profile remains largely unchanged, indicating that our physical interpretation is robust. 
    }
    \label{fig:conv_test}
\end{figure}

In Figure~\ref{fig:conv_test}, we show the outcome of our sensitivity analysis under different mass resolutions and box sizes of the simulations considered. We report the results relative to $z=0$ only, having verified that we reach qualitatively similar conclusions at higher redshifts. The two columns of panels refer to the first-order and total-order Sobol indices. We represent the former with data points of different marker styles and colours to distinguish different simulations. As in Figures~\ref{fig:sobol_z}-\ref{fig:sobol_simba}, the vertical error bars represent the 95\% confidence level of the estimates of $\mathbf{S}_1$. For clarity, we remove the horizontal error bars designating the radial bin width. We additionally shift the Simba and EAGLE datasets by 5\% with respect to the bin centre to improve the legibility of the figure, although all simulations share the same radial binning.

It is worth highlighting that Figure~\ref{fig:conv_test} does not constitute a convergence test of the gas density profiles or galaxy properties in EAGLE, IllustrisTNG, and Simba. The dynamic range of the input features can vary even for simulations with the same galaxy formation model, but different volumes or mass resolutions. Thus, the dataset on which our RF algorithm is trained changes for every run considered. Therefore, there is no reason to expect that the algorithm itself converges to a defined solution, for a given feedback scenario. Consequently, the associated Sobol indices are not a priori guaranteed to converge either. Nevertheless, we should test whether the rank ordering of the indices relative to the different input features is preserved when changing volume and mass resolution. If that is the case, then our physical interpretation of the sensitivity analysis undertaken in \S~\ref{sec:sensitivity}-\ref{sec:discussion} is robust. Testing this is the purpose of Figure~\ref{fig:conv_test}.

From a first inspection, we note that having a sufficiently large volume is crucial for obtaining well-converged Sobol indices. In EAGLE (Simba), the results for the $50 \cMpc$ and $100 \cMpc$ ($75 \cMpc$ and $150 \cMpc$) boxes are largely consistent with each other. This indicates that a box size of at least $50 \cMpc$ is needed to capture a high enough statistic of haloes and density profiles and ensure convergence in the corresponding Sobol indices. In the case of IllustrisTNG, we always combine all boxes from $50 \cMpc$ to $300 \cMpc$ together, which leads to a higher level of volume-wise convergence by construction. 

All Sobol indices are well converged also with respect to mass resolution for the IllustrisTNG simulations. Instead, the Simba-25 and Simba-25-HighRes runs exhibit statistically significant differences in the Sobol indices for most features. This is not too surprising, since the feedback parameters of the higher-resolution simulation are not re-calibrated to match the galaxy stellar mass function as in the Simba-25 run. Therefore, even though the two simulations start from the same initial conditions and contain the same physical model, they are not expected to achieve full convergence in the underlying galaxy and halo properties.

We find poor robustness with respect to mass resolution for the EAGLE simulation too, despite the volume being larger ($100 \cMpc$). No matter how large the simulation volume, the halo sample will still be dominated by lower-mass haloes due to the steepness of the halo mass function. As it happens, EAGLE exhibits an intrinsically larger variance in the gas density profiles at the lower-mass end compared with the other simulations in this work (see Figure~\ref{fig:dens_prof}). This may explain why resolution-wise convergence of the Sobol indices is more challenging to achieve for EAGLE.

A general point to be made for all simulations is that when we increase the mass resolution we automatically bias the halo sample towards lower masses, since we set a floor to the total number of particles of a halo upon including it in our training set (see \S~\ref{sec:profiles}). Key properties such as SFR or BH mass diversely impact haloes in different mass ranges due to the prevalence of distinct stellar or AGN feedback modes. Because the physics is intrinsically different when changing the dynamic range of $M_{\rm 200c}$, it is not surprising that we find different results for the subsequent sensitivity analysis when switching to boxes with higher mass resolution. Thus, the approach that we adopted for the IllustrisTNG simulations, i.e. combining boxes with different volumes and mass resolutions, seems to be the most promising to construct robust data-driven models.
 
Nevertheless, the most important takeaway is that even when there is disagreement in the {\it magnitude} of the Sobol indices with respect to mass resolution (EAGLE and Simba), the {\it relative rank ordering} of the importance of different features for each run is mostly unchanged. This again underscores the robust physical interpretability of our model. 

\vspace{10ex}


\end{document}